\documentclass[manuscript]{emulateapj}
\usepackage{amssymb}
\usepackage[export]{adjustbox}
\usepackage{amsmath}
\usepackage{booktabs}
\usepackage{hyperref}
\usepackage{multirow}
\usepackage{times}
\PassOptionsToPackage{hyphens}{url}\usepackage{hyperref}
\usepackage[hyphenbreaks]{breakurl}

\newcommand{\mcg}{MCG$-$03$-$58$-$007}
\newcommand{\degg}{\hbox{$^\circ$}}
\newcommand{\zw}{I\,Zw\,1}
\newcommand{\pg}{PG\,1448+273}
\newcommand{\chinu}{\chi^2/\nu}
\newcommand{\xmm}{{\it XMM-Newton}}

\newcommand{\arcs}{\hbox{$^{\prime\prime}$}}

\newcommand{\ls}
{\mathrel{\hbox{\rlap{\hbox{\lower4pt\hbox{$\sim$}}}\hbox{$<$}}}}
\newcommand{\gs}
{\mathrel{\hbox{\rlap{\hbox{\lower4pt\hbox{$\sim$}}}\hbox{$>$}}}}

\received{}
\begin{document}
\title{An Eddington Limited Accretion Disk Wind in the narrow line Seyfert 1, PG\,1448+273.}
\shorttitle{A Variable Disk Wind in PG\,1448+273.}
\shortauthors{Reeves et al.}
\author{J. N. Reeves\altaffilmark{1,2}, V. Braito\altaffilmark{1,2,3}, A. Luminari\altaffilmark{4,5}, D. Porquet\altaffilmark{6}, M. Laurenti\altaffilmark{7,8}, G. Matzeu\altaffilmark{9}, A. Lobban\altaffilmark{10}, S. Hagen\altaffilmark{11}}
\altaffiltext{1}{Department of Physics, Institute for Astrophysics and Computational Sciences, The Catholic University of America, Washington, DC 20064, USA, email reevesjn@cua.edu}
\altaffiltext{2}{INAF, Osservatorio Astronomico di Brera, Via Bianchi 46 I-23807 Merate (LC), Italy}
\altaffiltext{3}{Dipartimento di Fisica, Universit\`a di Trento, Via Sommarive 14, Trento 38123, Italy.}
\altaffiltext{4}{INAF - IAPS, via del Fosso del Cavaliere 100, 00100 Roma}
\altaffiltext{5}{INAF–Osservatorio Astronomico di Roma, Via Frascati33, 00078 Monteporzio, Italy}
\altaffiltext{6}{Aix-Marseille Univ., CNRS, CNES, LAM, Marseille, France}
\altaffiltext{7}{Dipartimento di Fisica, Universit\`a di Roma “Tor Vergata”, Via della Ricerca Scientifica 1, 00133 Roma, Italy}
\altaffiltext{8}{Space Science Data Center, SSDC, ASI, Via del Politecnico snc, 00133 Roma, Italy}
\altaffiltext{9}{Quasar Science Resources SL for ESA, European Space Astronomy Centre (ESAC), Science Operations Department, 28692, Villanueva de la Ca\~{n}ada, Madrid, Spain}
\altaffiltext{10}{European Space Astronomy Centre (ESA/ESAC), E-28691 Villanueva de la Canada, Madrid, Spain}
\altaffiltext{11}{Centre for Extragalactic Astronomy, Department of Physics, University of Durham, South Road, Durham DH1 3LE, UK}


\begin{abstract}

\pg\ is a luminous, nearby ($z=0.0645$), narrow line Seyfert 1 galaxy, which likely accretes close to the Eddington limit. 
Previous X-ray observations of \pg\ with \xmm\ in 2017 and {\it NuSTAR} in 2022 revealed the presence of 
an ultra fast outflow, as seen through its blueshifted iron K absorption profile, where the outflow velocity 
appeared to vary in the range $0.1-0.3c$. In this work, new X-ray observations of \pg\ 
are presented, in the form of four simultaneous \xmm\ and {\it NuSTAR} observations performed in July and August 2023. 
The X-ray spectra appeared at a similar flux in each observation, making it possible to analyze the mean 2023 X-ray 
spectrum at high signal to noise. A broad ($\sigma=1$\,keV) and highly blue-shifted ($E=9.8\pm0.4$\,keV) iron K absorption profile 
is revealed in the mean spectrum. The profile can be modeled by a fast, geometrically thick accretion disk wind, 
which reveals a maximum terminal velocity of $v_{\infty}=-0.43\pm0.03c$, one of the fastest known winds in a nearby AGN. 
As a result, the inferred mass outflow rate of the wind may reach a significant fraction of the Eddington accretion rate. 
\end{abstract}










\section{Introduction}

Since their initial discovery two decades ago \citep{Chartas02, Pounds03, Reeves03}, ultra fast outflows (or UFOs) in X-rays have been found to occur 
in about 40\% of Active Galactic Nuclei \citep{Tombesi10}. Most of these detections arise through studies of the iron K band 
towards nearby and X-ray bright, local ($z<0.1$) Seyfert~1 AGN, with typical outflows velocities of around $0.1c$; e.g. \citealt{Tombesi10, Gofford13, Igo20, Matzeu23}. 
The mechanical power of these winds can typically reach up to a few percent of the Eddington limit \citep{Tombesi13, Gofford15}, sufficient to provide significant mechanical feedback into the AGN host galaxy (\citealt{SilkRees98, Fabian99,DiMatteo05,King03,HopkinsElvis10}). 
Powerful black hole winds may also play a crucial part in the regulating the growth of super massive black holes in more luminous quasars  \citep{Nardini15, Tombesi15, Matzeu23} and 
notably at high redshifts near the peak of the QSO evolution (e.g. \citealt{Chartas21} and references therein). 

At lower redshifts, the Narrow Line Seyfert 1s (NLS1s, \citealt{OP85}) represent a part of the AGN population with relatively low black hole masses, yet with high 
accretion rates with respect to the Eddington limit \citep{Boroson02}. As such they provide an excellent laboratory for studying accretion disk winds, which produce the observed 
ultra fast outflows \citep{PK04, Sim08, Sim10a, Fukumura10, Mizumoto21}.  Several notable examples of ultra fast outflows have been detected 
in nearby NLS1s \citep{PoundsVaughan12, Longinotti15, Hagino16, Parker17, Kosec18, ReevesBraito19}, expanding the parameter space of AGN accretion disk winds into the 
high accretion rate regime. 

The subject of this paper is the nearby ($z=0.0645$) NLS1, \pg, which also shows evidence for a powerful UFO, observed 
both in the iron K band and at soft X-rays \citep{Kosec20, Laurenti21, Reeves23}.  
It is also classed as a radio-quiet QSO \citep{SchmidtGreen83} and has narrow permitted lines, e.g. 
H$\beta$ FWHM of 1330\,km\,s$^{-1}$ \citep{Grupe04}, with a bolometric luminosity estimated to be $L_{\rm bol}=2-3\times10^{45}$\,erg\,s$^{-1}$ (e.g., \citealt{Grupe04, Rakshit20}). 
Recently, a black hole mass estimate of $M_{\rm BH}=1.01^{+0.38}_{-0.23}\times10^{7}$\,M$_{\odot}$ was derived by \citet{Hu21}, obtained from a H$\beta$ based reverberation study of 15 PG quasars. Given its observed bolometric luminosity, this implies that \pg\ is likely to accrete near the Eddington limit, which may provide favorable conditions for launching an accretion disk wind, as discussed by \citet{GP19}. 
 

\begin{deluxetable*}{lccccccc}
\tablecaption{Observation log for the 2023 \xmm, {\it NuSTAR} and {\it Swift} Campaign of \pg.}
\tablewidth{500pt}
\tablehead{
\colhead{Observation} & \colhead{Sequence} & \colhead{Start Time (UT)} &  \colhead{End Time (UT)} & \colhead{Duration (ks)} & \colhead{Net Exp (ks)$^a$} & \colhead{Rate(s$^{-1}$)$^b$} &  \colhead{Flux$^c$}}
\startdata
NuSTAR 1 & 6092001002 & 2023/07/05 22:16:09 & 2023/07/07 03:06:09 & 103.8 & 51.7 & $0.074\pm0.002$ & 3.2\\
NuSTAR 2 & 6092001004 & 2023/07/17 18:36:09 & 2023/07/19 00:56:09 & 109.2 & 51.1 & $0.069\pm0.002$ & 3.0\\
NuSTAR 3 & 6092001006 & 2023/08/02 16:51:09 & 2023/08/03 23:06:09 & 108.9 & 52.4 & $0.062\pm0.002$ & 2.7\\
NuSTAR 4 & 6092001008 & 2023/08/17 10:11:09 & 2023/08/18 15:01:09 & 103.8 & 54.2 & $0.063\pm0.002$ & 2.8\\
NuSTAR mean & & & & & 196.1$^d$ & $0.069\pm0.001$ & 3.0\\
\hline\\
XMM-Newton 1 & 0920250201 & 2023/07/06 00:58:34 & 2023/07/07 00:53:57 & 86.1 & 80.3 & $3.555\pm0.007$ & 5.9\\
XMM-Newton 2a & 0920250301 & 2023/07/18 00:09:33 & 2023/07/18 03:25:52 & 11.8 & 6.3 & $2.108\pm0.022$ & 3.7\\
XMM-Newton 2b & 0920250301 & 2023/07/18 08:33:29 & 2023/07/18 21:18:44 & 45.9 & 34.3 & $3.646\pm0.010$ & 5.9\\
XMM-Newton 3 & 0920250401 & 2023/08/02 18:11:32 & 2023/08/03 16:31:34 & 73.2 & 53.4 & $3.216\pm0.008$ & 5.1\\
XMM-Newton 4 & 0920250601 & 2023/08/17 09:55:40 & 2023/08/17 16:04:02 & 22.1 & 20.6 & $3.842\pm0.014$ & 6.1\\
XMM-Newton mean & & & & & 188.6$^d$ & $3.481\pm0.004$ & 5.8\\
\hline\\
Swift/XRT (obs 1--39)$^e$ & 00097192 & 2023/06/25 10:12:35 & 2023/08/14 11:08:56 & & 35.5 (total) & $0.071\pm0.008$ & $2.0$ (low) \\
& & & & & & $0.389\pm0.028$ & 12.3 (high)\\
& & & & & & $0.181\pm0.013$ & 5.3 (mean)\\
\enddata
\tablenotetext{a}{Net exposure, for \xmm\ (EPIC-pn), NuSTAR and Swift/XRT, correcting for background screening, SAA passage and detector deadtime.}
\tablenotetext{b}{Net count rates per \xmm\ (EPIC-pn), Swift/XRT or {\it NuSTAR} observation.}
\tablenotetext{c}{Observed flux measured from 0.5-10\,keV for \xmm\ (EPIC-pn) and Swift/XRT and 3-30\,keV band for {\it NuSTAR}. Units are $\times10^{-12}$\,ergs\,cm$^{-2}$\,s$^{-1}$.}
\tablenotetext{d}{The total exposure time of the mean spectrum excludes the short lower flux OBS\,2a \xmm\ sequence and the corresponding dip portion of the 2nd {\it NuSTAR} observation, from 20--50\,ks as measured from the start of that observation (see Figure~\ref{fig:xmmlc}).}
\tablenotetext{e}{The minimum, maximum and mean count rates and fluxes are given across all 39 {\it Swift} XRT observations.}
\label{tab:obs}
\end{deluxetable*}

  
The detection of a fast wind in \pg\ was initially obtained by \citet{Kosec20} and \citet{Laurenti21} (see their Figures~3 and 1 respectively), on the basis of a 75\,ks \xmm\ exposure in 2017. 
This 2017 \xmm\ observation occurred just prior to a pronounced dip as observed by the {\it Swift} monitoring \citep{Laurenti21}, resulting in an historically low X-ray flux for \pg. 
An Fe K absorption trough was measured centered at 7.5\,keV, which implied an outflow velocity of $\sim0.1c$, assuming an identification with the $1s\rightarrow2p$ resonance lines from He-like (Fe\,\textsc{xxv}) and 
H-like (Fe\,\textsc{xxvi}) iron. It has one of the deepest absorption profiles amongst the UFOs reported to date (c.f. \citealt{Tombesi10, Gofford13}), with an equivalent width of  ${\rm EW}=-410\pm80$\,eV. 
As was inferred by \cite{Kosec20} and \citet{Laurenti21}, the high equivalent width implies an absorption column density of the order $N_{\rm H}=10^{24}$\,cm$^{-2}$. In particular, the whole absorption profile studied by \cite{Laurenti21} was modeled with a physically motivated accretion disk wind model; utilizing the \textsc{wine} (Wind in the Ionised Nuclear Environment) code of \citet{Luminari18}.

The first {\it NuSTAR} observation of \pg\ (250\,ks duration, 125\,ks exposure) occurred in January 2022 and revealed strong X-ray variability \citep{Reeves23}. 
The first part of the observation (slice A), seen at a much higher flux ($F_{2-10}=4.8\times10^{-12}$\,erg\,cm$^{-2}$\,s$^{-1}$), was coincident with a 70\,ks {\it XMM} exposure 
and the wind absorption at Fe K had diminished in opacity compared to 2017. During the last 60\,ks with {\it NuSTAR} (slice B), a deep ($>\times2$) and rapid ($<10$\,ks) drop in flux was observed and was also accompanied by an increase in spectral hardness.  
The slice B {\it NuSTAR} spectrum revealed a deep Fe K absorption trough at 9\,keV and could be modeled by a disk wind model \citep{Sim08, Sim10a, Matzeu22} of terminal velocity $v_{\infty}=-0.26\pm0.03c$, 
which implies at least a factor of $\times2$ increase 
in wind velocity compared to the original 2017 observation. Thus the wind in \pg\ appears to be strongly variable, both in its opacity and velocity, reminiscent of the drastic velocity changes in MCG--3--58--07 
($v/c\sim0.07\rightarrow0.2$; \citealt{Braito22}).  

This paper presents a follow-up of the \citet{Reeves23} work, where subsequently four new simultaneous \xmm\ and {\it NuSTAR} observations of \pg\ were performed in July and August 2023. The observations were also co-ordinated with daily {\it Swift} monitoring, to measure the overall variability of \pg\ throughout the campaign. In these new observations, the iron K wind profile is found to be both velocity broadened and blue-shifted up to 10\,keV in the X-ray spectra, revealing one of the fastest known ultra fast outflows (UFOs) in an AGN X-ray spectrum. The overall profile can be modeled by a geometrically thick disk wind, which 
achieves a maximum terminal velocity of $v_{\infty}=-0.43\pm0.03c$. The subsequent mass outflow rate is inferred to be close to the Eddington rate in this high accretion rate AGN. 

The paper is organized as follows. 
In Section~2, the observations and data reduction are described, 
while in Section~3 the AGN variability is quantified through the \xmm, {\it NuSTAR} and {\it Swift} observations. Section~4 presents the results of the physically motivated disk wind modeling, where the velocity range and mass 
outflow rate of the AGN wind are inferred. Section~5 quantifies any wind variability within the 2023 campaign and is compared with past observations. Section~6 then compares the properties of the wind in \pg\ with other AGN, while the wind energetics and plausible launching mechanisms are discussed. 
Throughout the paper, 90\% confidence intervals for one parameter of interested are adopted for the uncertainties (or $\Delta \chi^2=2.7$), while parameters are stated in the AGN rest frame 
at $z=0.0645$. The standard $\Lambda$CDM cosmology ($H_0 = 70$\,km\,s${-1}$\, Mpc\,$^{-1}$, $\Omega_m = 0.3$, $\Omega_{\Lambda} = 0.7$) is adopted throughout the paper.

\section{Observations and Data Reduction}

The 2023 X-ray campaign of \pg\ encompassed four observations with \xmm\ \citep{Jansen01} and {\it NuSTAR} \citep{Harrison13} in July--August 2023, with a spacing of about two weeks between each observation; see Table~\ref{tab:obs} for the observation log of the campaign. 
The {\it Swift} 
campaign consisted of 39 daily X-ray Telescope (XRT, \citealt{Burrows05}) snapshots of approximate 1\,ks duration from June--August 2023, in order to provide temporal coverage over the entire \xmm\ and {\it NuSTAR} campaign. The {\it Swift} XRT data were processed with v0.13.7 of the \textsc{xrtpipeline} to create the lightcurves and spectra. A circular source extraction region of 20\arcs\ was used, while for the background 
an annulus of radii 40\arcs\ and 130\arcs\ centered on the source was adopted. 

The first and third of the  \xmm\ observations (hereafter OBS\,1 and OBS\,3) covered a duration of about a day (86.1\,ks and 73.2\,ks respectively), as measured from the start and stop times of the sequences (see Table~\ref{tab:obs}). The second \xmm\ observation (OBS\,2) was interrupted by a strong Solar flare of 5 hours duration and the telescope filter wheel was subsequently closed during this time (\textsc{cal closed} position). 
As a result, the second \xmm\ observation was split into two separate sequences either side of the Solar flare (hereafter OBS\,2a and 2b), of duration 11.8\,ks and 45.9\,ks respectively. 
The fourth \xmm\ observation (OBS\,4) was scheduled as a result of the exposure time lost during the second observation and was of shorter (22.1\,ks) 
duration. 
Each of the \xmm\ observations were performed simultaneously with {\it NuSTAR} at hard X-rays. 
All four {\it NuSTAR} observations covered a total duration of about 100\,ks each, in order to overlap with the start and stop times of each of the \xmm\ exposures, while the 
fourth \xmm\ observation coincided with the start of the corresponding {\it NuSTAR} exposure (see Table~\ref{tab:obs}).
The \xmm\ EPIC-pn \citep{Struder01} exposures were performed in Large Window mode, with the medium filter applied, while the EPIC-MOS \citep{Turner01} exposures were in Small Window mode. 

\begin{figure}
\begin{center}
\hspace*{-0.8cm}
\rotatebox{-90}{\includegraphics[height=8.7cm]{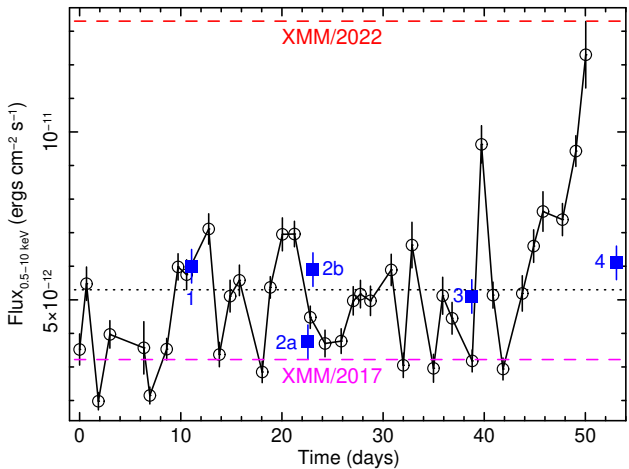}}
\end{center}
\vspace*{-0.3cm}
\caption{{\it Swift} XRT lightcurve of \pg\ from the 2023 monitoring campaign, where time is measured in days from the start time of the {\it Swift} monitoring 
listed in Table~\ref{tab:obs}. The 0.5--10\,keV XRT lightcurve (black circles) exhibits a factor of $\times 6$ variability over the whole campaign.
In comparison, the 0.5--10\,keV flux measured from the 4 \xmm\ observations (blue squares) show little variability between each sequence, with the exception of the 
short OBS\,2a sequence. Overall the \xmm\ observations captured the source close to the average {\it Swift} flux in the campaign, as is shown by the black dotted horizontal line. In contrast, two previous \xmm\ observations of \pg\ in 2017 and 2022 spanned the range of {\it Swift} flux, as is indicated by the magenta and red horizontal dashed lines. 
Thus the 2023 \xmm\ observations resemble an average flux state of \pg.}
\label{fig:swift} 
\end{figure}

The \xmm\ and {\it NuSTAR} observations were processed using the \textsc{nustardas} v2.1.2, 
{\it XMM-Newton} \textsc{sas} v20.0 and \textsc{heasoft} v6.30 software. 
Background screening was performed on the \xmm\ observations, in order to remove periods of high background due to Solar flares during the course of each \xmm\ orbit. 
A background cut of $>1$\,cts\,s$^{-1}$ from 10--12\,keV over the 
entire EPIC-pn field of view was applied to select good time intervals of low background for subsequent spectral extraction. While the first and fourth observations were free of background flares, a strong Solar flare affected \xmm\ OBS\,2 (as noted above). As a result, the net EPIC-pn exposures were reduced to 6.3\,ks and 34.3\,ks for the OBS\,2a and OBS\,2b sequences 
respectively. A portion of \xmm\ OBS\,3 was also affected by background flares, which resulted in a reduction of the net exposure to 53.4\,ks for spectral extraction. 
The {\it NuSTAR} exposures were screened for passage through the South Atlantic Anomaly (SAA), resulting in net exposures of between 51--54\,ks per FPM module per observation; this is typical of the 50\% observing efficiency in a low Earth orbit. Table~\ref{tab:obs} summarizes all of the resulting exposures.

{\it NuSTAR} source spectra were extracted using a 45\arcs\ circular region centered on the source and two background  circular regions with  a 45\arcs\  radius and clear from stray light. 
{\it XMM-Newton} EPIC-pn spectra were extracted from single and double events (patterns 0--4), using a 35\arcs\ source region and $2\times35\arcs$\ background regions on the same chip. 
EPIC-MOS spectra were extracted using patterns 0--12, using a 30\arcs\ source region and $2\times36\arcs$\ background regions. The effective area of the EPIC CCD detectors was 
corrected at high energies through the \textsc{sas} task \textsc{arfgen} by applying the option \textsc{applyabsfluxcorr}\footnote{https://xmmweb.esac.esa.int/docs/documents/CAL-SRN-0388-1-4.pdf}. This option provides an improved cross calibration for simultaneous observations between {\it NuSTAR} and \xmm\ over the overlapping 3--10\,keV range. 
For each observation, the spectra and responses from the individual FPMA and FPMB detectors on-board {\it NuSTAR} 
were combined into a single spectrum after they were first checked for consistency. The {\it NuSTAR} spectra were utilized over the 3--30\,keV band; above 30\,keV the source spectrum becomes background dominated as the source count rate declines. The total {\it NuSTAR} background count rate over this band is approximately 6\% of the source rate. For EPIC-pn the background rate after filtering is about 4\% over the 3--10\,keV band and negligible at soft X-rays. 

As the effective area of the EPIC-pn detector is about a factor of $\times8$ larger than that of each EPIC-MOS CCD at 9 keV, where the high energy absorption feature in \pg\ occurs (see Section 4), 
we used only the EPIC-pn and {\it NuSTAR} data in the detailed spectral analysis. Nonetheless the MOS spectra are found to be consistent with the EPIC-pn across the 0.3--10\,keV band. 
All the spectra are binned to a minimum of 100 counts per bin to ensure a minimum signal to noise ratio of 10 in the spectra and $\chi^2$ minimization was used for the subsequent spectral fitting. 
Count rates and fluxes for each of the exposures are listed in Table~\ref{tab:obs}.

Spectra from the {\it XMM-Newton} Reflection Grating Spectrometer (RGS, \citealt{denHerder01}) for all of the 2023 observations were extracted using the \textsc{rgsproc} pipeline. These were combined into a single RGS\,1+2 spectrum for each observation using using \textsc{rgscombine}, after first checking that the individual 
RGS 1 and RGS 2 spectra were consistent with each other within the errors. Furthermore, very little variability was observed over the 4 epochs (OBS\,1, 2b, 3, 4) in the RGS band; the 
0.4--2.0\,keV flux ranged between $3.6-4.1\times10^{-12}$\,ergs\,cm$^{-2}$\,s$^{-1}$, while no spectral variations were seen. As a result, all of these epochs were combined into a single 
2023 RGS spectrum to maximize signal to noise, with the exception of the very short OBS\,2a sequence, which has very low S/N. 
The total net count rate obtained over the 6--30\,\AA\ band (RGS\,1+2 combined) was $0.195\pm0.001$\,cts\,s$^{-1}$, with a net exposure of 141.7\,ks. 
The spectra were binned in constant wavelength bins of $\Delta\lambda=0.1$\,\AA, which approximates the spectral resolution of the RGS gratings. 

\begin{figure}
\begin{center}
\hspace*{-0.5cm}
\rotatebox{-90}{\includegraphics[height=9.5cm]{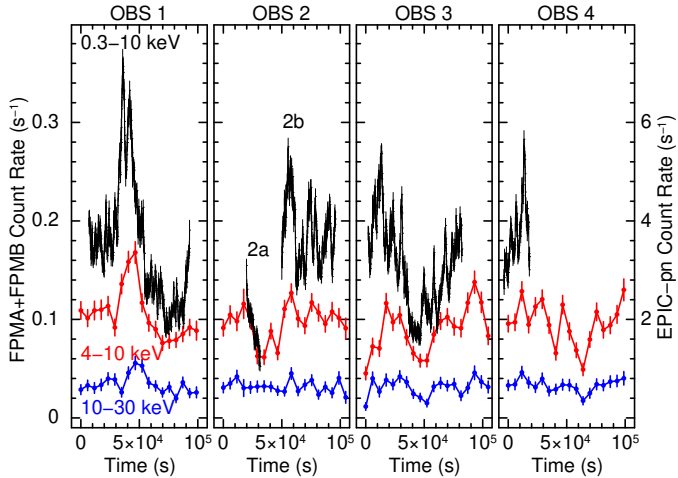}}
\end{center}
\vspace*{-0.3cm}
\caption{Background subtracted lightcurves from the 2023 \xmm\ and {\it NuSTAR} observations of \pg. The x-axis shows the time elapsed for each observation, compared to the start time of 
each {\it NuSTAR} exposure, while the y-axis shows the count rates for {\it NuSTAR} (FPMA+FPMB) and EPIC-pn (right-hand axis). 
The 0.3--10\,keV EPIC-pn lightcurve is shown in black in 200\,s bins, while {\it NuSTAR} is shown in red and blue, over the 4--10\,keV and 10--30\,keV bands respectively and are binned into 5814\,s orbital bins. 
Note the gap in \xmm\ OBS\,2, split into 2 sequences (2a and 2b as marked), was due to the Solar flare which interrupted the observation. Overall the average source flux was similar between each of the observations, while a notable flare was observed in OBS\,1 and OBS\,2a appears to coincide with a short dip in the lightcurve. 
The variability amplitude in the hardest 10--30\,keV band is modest in comparison.}
\label{fig:xmmlc} 
\end{figure}

\section{Overall AGN Variability}


Figure~\ref{fig:swift} shows the fluxed {\it Swift} XRT lightcurve of \pg\ during the 2023 campaign. Here, the AGN displayed a factor of $\times6$ variability, ranging in 0.5--10\,keV flux from $2-12\times10^{-12}$\,ergs\,cm$^{-2}$\,s$^{-1}$. Several mini X-ray flares are apparent in the lightcurve, the most notable occurring at the end of the {\it Swift} monitoring.  For comparison, the mean flux of each of the \xmm\ observations are overlaid on the Figure~\ref{fig:swift} lightcurve as blue squares. In contrast to the variability amplitude caught by {\it Swift}, each of the \xmm\ observations (OBS\,1, 2b, 3, 4) caught the source at a very similar flux, covering a narrow range 
from $F_{0.5-10\,{\rm keV}}=5.1-6.1\times10^{-12}$\,ergs\,cm$^{-2}$\,s$^{-1}$ and close to the mean {\it Swift} flux of $F_{0.5-10\,{\rm keV}}=5.3\times10^{-12}$\,ergs\,cm$^{-2}$\,s$^{-1}$. 
The only exception to this occurs at the very start of second \xmm\ observation (OBS\,2a sequence), where the AGN was seen at a
lower flux of $F_{0.5-10\,{\rm keV}}=3.7\times10^{-12}$\,ergs\,cm$^{-2}$\,s$^{-1}$. Even the fourth \xmm\ observation caught the AGN close to the average {\it Swift} flux, despite it occurring a few days after the strong X-ray flare at the end of the {\it Swift} lightcurve. 
The lack of any strong variability comparing each of the 2023 \xmm\ observations also contrasts with two recent \xmm\ observations of \pg\ in 2017 and 2022, which encompass the range of fluxes 
observed in the {\it Swift} lightcurve, see Figure~\ref{fig:swift}. In contrast, the four 2023 \xmm\ observations appear to probe an average flux state of this AGN. 


X-ray lightcurves were also extracted for each of the observations over the full 0.3--10\,keV band for EPIC-pn, using time bins of 200\,s and over the 4--10\,keV and 10--30\,keV bands for {\it NuSTAR}, using orbital bins of 5814\,s duration; these are plotted in Figure~\ref{fig:xmmlc}. 
As can be seen from the lightcurves, the mean fluxes across all of the observations are similar, while factor of two variability on shorter kiloseconds timescales is observed. In particular, a strong flare is observed in OBS\,1, which is apparent across both the \xmm\ and {\it NuSTAR} lightcurves, while the start of OBS\,2a coincides with a factor of two dip in flux. Unfortunately, the recovery 
from this dip is missed by \xmm\ as it occurs during the \textsc{cal closed} segment of OBS\,2, although the 4--10\,keV band {\it NuSTAR} lightcurve does capture the dip from 20--50\,ks in its entirety. 
The highest energy band (10--30\,keV) exhibits little variability, indicating the intrinsic hard X-ray continuum remained relatively steady over the entire campaign. 

\begin{figure}
\begin{center}
\hspace*{-0.8cm}
\rotatebox{-90}{\includegraphics[width=7cm]{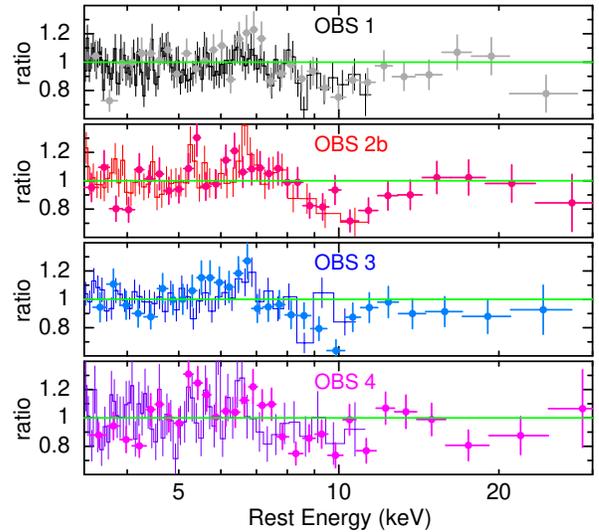}}
\end{center}
\vspace*{-0.3cm}
\caption{The four panels shows the simultaneous \xmm\ and {\it NuSTAR} (filled circles) spectra for OBS\,1, 2b, 3 and 4 
(top to bottom), plotted as a ratio to a power-law over the 3--30\,keV band. Note the deficit of counts observed from 8--12\,keV in each case, highlighting the 
presence of an absorption trough.}
\label{fig:fourobs}
\end{figure}

\section{X-ray Spectral Analysis}

A simple comparison of the individual spectra was then performed across all four of the \xmm\ observations, excluding the very short OBS\,2a dip exposure. 
The \xmm\ spectra were fitted with a simple power-law model modified by Galactic absorption, along with their corresponding simultaneous {\it NuSTAR} spectra over the 3--30\,keV band. 
Figure~\ref{fig:fourobs} shows the spectra plotted as a data-to-model ratio to the above power-law model. 
None of the spectra show a significant variation in photon index within $\Delta\Gamma=0.1$ (e.g., OBS\,1, $\Gamma=2.17\pm0.03$ vs OBS\,4, $\Gamma=2.23\pm0.05$).  
The hard X-ray flux, as measured by {\it NuSTAR}, also shows little variation 
($F_{3-30 \rm{keV}}=2.7-3.1\times10^{-12}$\,ergs\,cm$^{-2}$\,s$^{-1}$). Negative residuals are also present in each of the ratio spectra between 8--12\,keV, indicating the presence 
of iron K absorption and positive residuals are observed between 5--7\,keV, indicating emission. 
Below 3\,keV, all of the spectra show a smooth soft X-ray excess above the power-law continuum, with very similar fluxes over the four observations (Table~\ref{tab:obs}). 

\begin{figure*}
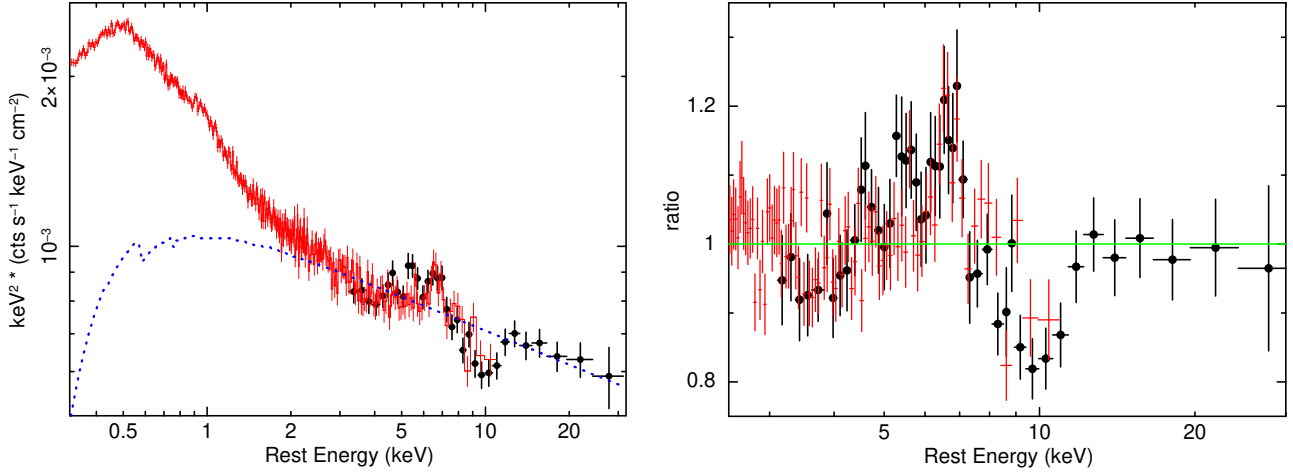

\begin{center}
\hspace*{-0.8cm}
\rotatebox{-90}{\includegraphics[height=8.7cm]{fig4a.eps}}
\rotatebox{-90}{\includegraphics[height=8.7cm]{fig4b.eps}}
\end{center}
\vspace*{-0.3cm}
\caption{Mean \xmm\ EPIC-pn and {\it NuSTAR} X-ray spectra of \pg. The pn is shown in red and {\it NuSTAR} FPMA+B as black circles. The left panel shows the fluxed spectrum, where the count rate spectrum has been divided by the instrumental effective area and multiplied twice by energy and thus the flux is in $\nu F_{\nu}$ units. The dotted blue line shows the approximate level of the hard X-ray power-law continuum, modified by Galactic absorption. 
The strong soft X-ray excess is apparent above the power-law. Note the cross normalization factor of $\times 1.14$ between {\it NuSTAR} and \xmm\ is accounted for in the plot. 
The right panel shows the data/model ratio to the baseline two component continuum model 
described in Section~4. 
Both plots highlight the strong absorption trough centered near 10\,keV in the AGN rest frame, while Fe K band emission is observed between 5--7\,keV.}
\label{fig:spectra}
\end{figure*}

The individual 3--30\,keV band spectra were fitted by adding a broadened Gaussian absorption profile to the power-law continuum to model the absorption trough between 8--12\,keV. 
This is well modeled at a rest frame energy of $E=9.8\pm0.4$\,keV, with a width of $\sigma=1.2\pm0.5$\,keV; this improved the fit statistic from 
$\chinu=200.4/131$ (power-law only) to $\chinu=147.2/125$ (with absorption profile).  
The absorption line equivalent width does not vary across the four 2023 epochs, ranging from ${\rm EW}=-465\pm180$\,eV (OBS\,1) to ${\rm EW}=-570\pm210$\,eV (OBS\,4); i.e. the values are consistent within the errors. 
Likewise, allowing the line energy to vary between observations did not improve the fit ($\Delta\chi^2<2$), where the energy varies by $<0.5$\,keV. Thus the 
iron K absorption profile does not appear to vary between observations. 

Thus given the lack of spectral variability across the 2023 observations, a mean spectrum was created for both the EPIC-pn and {\it NuSTAR} data, in order for the iron K region to 
be characterized at high signal to noise. 
The mean spectra include all time intervals, except for the short OBS\,2a \xmm\ sequence and the corresponding dip period between 20--50\,ks as seen 
in {\it NuSTAR} OBS\,2 (13.3\,ks net exposure). The net exposures of the mean spectra (after background screening), along with their count rates and fluxes are reported in Table~\ref{tab:obs}. 
The spectra are also plotted in Figure~\ref{fig:spectra} (left panel) which shows the mean fluxed spectra in $\nu F_{\nu}$ space, after multiplying twice by photon energy and are folded 
through the instrumental responses. The \xmm\ and {\it NuSTAR} spectra are in good agreement, allowing for a small cross normalization multiplicative factor between {\it NuSTAR} and EPIC-pn of 
$C=1.14\pm0.02$. Note that the background level lies well below the source spectra (by more than an order of magnitude), except for the 
highest energy {\it NuSTAR} bin at 30\,keV. The spectra clearly reveal the structure in the Fe K region against the steep continuum, with an excess of counts peaking below 6-7 keV and a broad deficit between 8--12\,keV, due to an absorption trough. The overall profile resembles the P Cygni like wind profile observed in PDS\,456 \citep{Nardini15}. 

A baseline model (model (a)) was constructed to fit the X-ray continuum between 0.3--30\,keV. This consisted of two continuum components; a steep ($\Gamma>2$) power-law to account for 
the hard X-ray emission and a Comptonized disk component (the \textsc{comptt} model within \textsc{xspec}, \citealt{Titarchuk94}) for the soft X-ray excess. Both of these are modified by the  Galactic absorption column 
of $N_{\rm H}=3\times10^{20}$\,cm$^{-2}$ \citep{Kalberla05} as modeled by the \textsc{tbabs} model \citep{Wilms00}. The best fit continuum parameters are reported in Table~\ref{tab:diskwind} (model (a)). In addition to the continuum form, a weak soft X-ray emission line is required in the pn spectrum, at an energy of $E=0.96\pm0.02$\,keV and an equivalent width 
of ${\rm EW}=5.8\pm1.3$\,eV. The emission line width is unresolved at the pn resolution, with $\sigma<0.07$\,keV. Its possible origin has been discussed previously in \citet{Reeves23} from 
the analysis of the 2022 RGS spectrum and it likely arises from photoionized emission over the Ne band from larger scale gas. 

Although the baseline model provides a good description of the soft X-ray band, the overall fit statistic is quite poor, with a reduced chi-squared of $\chinu=764/622$. 
Strong residuals arise over the 3--30\,keV band, in the form of a broad iron K emission and absorption profile (see Figure~\ref{fig:spectra}, right panel).
Indeed restricting the fit to just the 3--30\,keV band yields a reduced chi-squared of $\chinu=315.5/194$, rejected with a null hypothesis probability of $8\times10^{-8}$.  

To provide an initial non-physical parameterization of the iron K profile, both a Gaussian emission and absorption line were added to the baseline model. 
This returned a significant improvement in the fit statistic, to $\chinu=643.7/616$, while the fit statistic is also acceptable over the 3--30\,keV band ($\chinu=193.7/188$). 
A broadened emission line was required, at a centroid energy of $E=6.61\pm0.09$\,keV, with an equivalent width of ${\rm EW}=143\pm26$\,eV and 
a line width of $\sigma=0.30^{+0.16}_{-0.10}$\,keV. Interestingly, a very high centroid energy is found for the absorption profile, with $E=9.8\pm0.4$\,keV, which is 
significantly broadened ($\sigma=1.1^{+0.5}_{-0.4}$\,keV) and has a high equivalent width (${\rm EW}=-435^{+150}_{-190}$\,eV). The parameters are also consistent with those 
obtained in the individual 2023 sequences. 
The centroid energy implies a blue-shift of either $v/c\sim-0.33$ or $v/c\sim-0.36$ for an association with either H-like (at 6.97\,keV) or He-like iron (at 6.7\,keV) respectively. The degree of blue-shift may be amongst the most extreme of those measured in nearby ultra fast outflows to date \citep{Tombesi10, Gofford13, Igo20, Luminari21, Matzeu23} and is at the highest end of the velocity range inferred for PDS\,456 \citep{Matzeu17}.  

For a more physical parameterization of the absorber, the broad Gaussian absorption line was replaced by an \textsc{xstar} absorption model \citep{Kallman04}, using the same absorption grids used in the \citet{Reeves23} paper on the 2022 datasets. 
A large turbulence velocity is used, with $v_{\rm turb}=25000$\,km\,s$^{-1}$, to parameterize the breadth of the profile. The best fit parameters of the \textsc{xstar} model are:- $N_{\rm H}=6.2^{+1.3}_{-2.1}\times10^{23}$\,cm$^{-2}$, 
$\log\xi=5.27^{+0.19}_{-0.13}$ and outflow velocity $v/c=-0.34\pm0.02$. Note the outflow velocity here corresponds to the point of maximum opacity in the profile, with an effective dispersion on the red and blue wings of $\pm0.08c$ according to the turbulence velocity. The fit is statistically equivalent to the Gaussian case, with $\chinu=649.9/616$, while the inferred velocity is also consistent. 
The parameters are in broad agreement with those obtained by \citet{Reeves23} for the lower flux slice B 
interval of the 2022 {\it NuSTAR} observation.

Note in comparison to the broad ionized emission and absorption profile, only an upper-limit can be placed on any narrow 6.4\,keV Fe K$\alpha$ fluorescence line, with an equivalent width of $<30$\,eV. This is consistent with the X-ray Baldwin effect in higher luminosity or higher accretion rate AGN \citep{IT93, Bianchi07}

\begin{figure}
\begin{center}
\rotatebox{0}{\includegraphics[width=8.5cm]{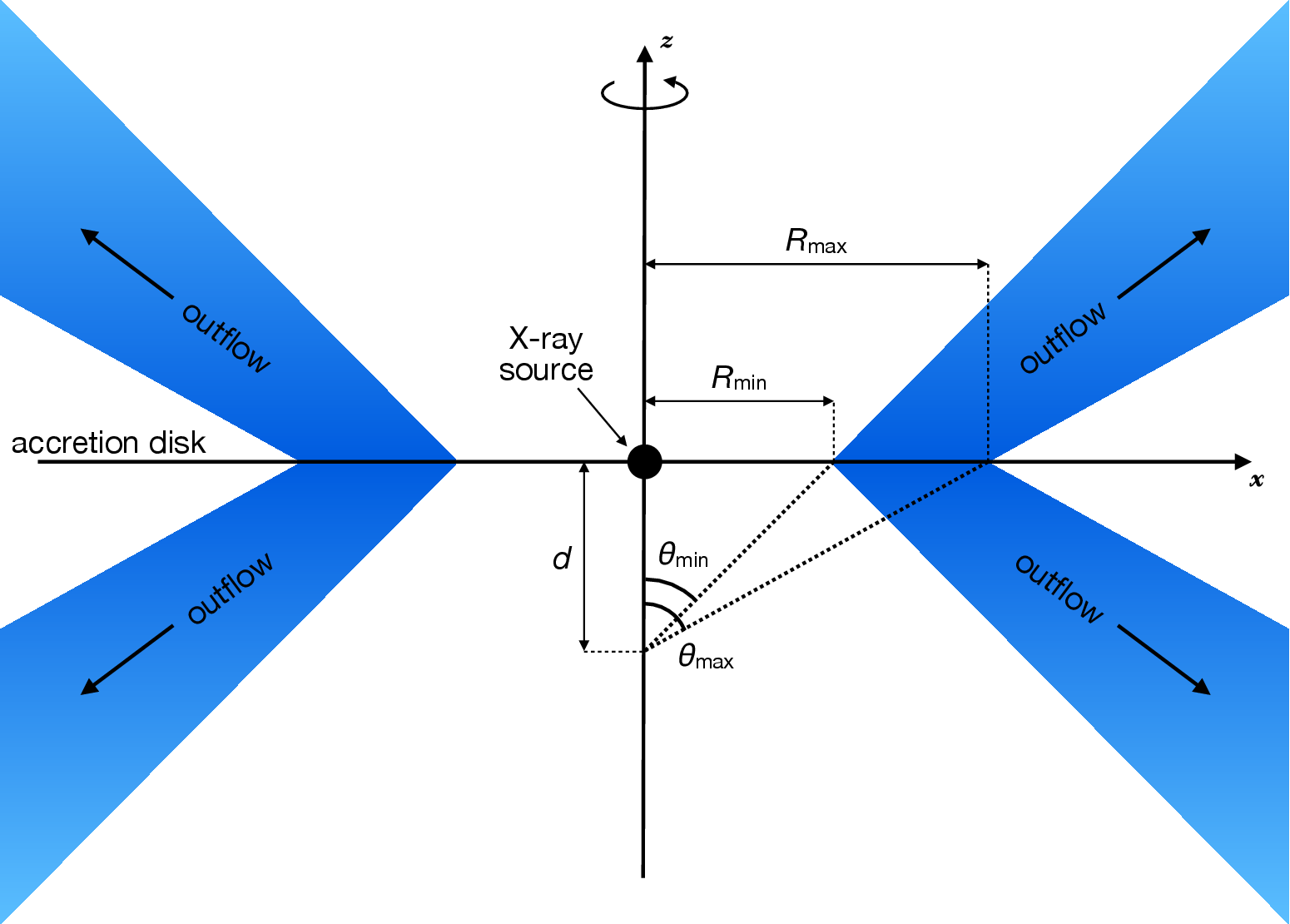}}
\end{center}
\vspace*{0.3cm}
\caption{Schematic of the disk wind geometry, adapted from \citet{Matzeu22}. The polar (z) and accretion disk axis (x) are shown as solid lines. The inner launching radius on the disk is $R_{\rm min}$ and the outer launching radius is $R_{\rm max}$. In the diskwind models adopted here, the minimum launch radius is set to $32R_{\rm g}$ (in units of the gravitational radius). 
The outer launching radius is $R_{\rm max}=48R_{\rm g}$ for the DW thin grid, while for the DW thick case $R_{\rm max}=96 R_{\rm g}$. Thus the latter scenario corresponds to a geometrically thicker wind. The minimum opening angle of the wind streamline is $\theta_{\rm min}$ and is given by $\arctan(R_{\rm min}/d)$. In the models explored here, $R_{\rm min}/d=1$ and thus $\theta_{\rm min}=45\degg$. See \citet{Matzeu22} and \citet{Sim08, Sim10a} for a more detailed model description.}
\label{fig:schematic} 
\end{figure}

\subsection{Disk Wind Modeling}

In order to model the Fe K wind signatures in the \pg\ spectra in a physical context, 
we utilized the radiative transfer disk wind code developed by \citet{Sim08,Sim10a,Sim10b}. 
The disk wind model provides a self consistent treatment of both the emission and absorption 
arising from a biconical wind, as well as computing the (non-uniform) ionization structure and velocity field 
through the flow. Photoionization and atomic data are adopted from \textsc{xstar} \citep{Kallman04}. 
The wind geometry is illustrated in Figure~\ref{fig:schematic}; see \citet{Matzeu22} for a more detailed description. 

This model has been previously employed to fit the Fe K wind absorption profiles in several AGN; e.g. Mrk 766 (\citealt{Sim08}), 
PG\,1211+143 (\citealt{Sim10a}), PDS\,456 (\citealt{Reeves14}), \zw\ \citep{ReevesBraito19}, MCG\,$-$03$-$58$-$007 \citep{Braito22}, PG\,1126$-$041 \citep{Giustini23}  
and most recently to the previous X-ray spectra of \pg\ \citep{Reeves23}. 
\citet{Matzeu22} expanded the parameter ranges covered by this wind model and tested the resulting grids on the 
prototype example of a fast disk wind in PDS\,456. The \textsc{fast32} grid calculated by \citet{Matzeu22} for this purpose was also recently applied to the previous (2017, 2022) 
X-ray spectra of \pg\ \citep{Reeves23}. 
Here, we test two variants of the \textsc{fast32} disk wind model:-

\begin{itemize}
\item A {\it geometrically thin} accretion disk wind (hereafter DW thin), which is identical to the \textsc{fast32} grid calculated by \citet{Matzeu22}. 
Here the inner launch radius is $R_{\rm min}=32R_{\rm g}$ (in gravitational units), while the outer launch radius on the disk surface is 
$R_{\rm max}=48R_{\rm g}$. 

\item A {\it geometrically thick} accretion disk wind (hereafter DW thick), where the inner launch radius is $R_{\rm min}=32R_{\rm g}$ and the outer launch radius is increased to  
$R_{\rm max}=96R_{\rm g}$. 
\end{itemize}

For clarity, we give the velocities and terminal velocities attained in the wind, as originally defined in \cite{Sim08, Sim10a}. 
Consider the velocity as a function of distance along the flow launched at a single point off the disk. This is expressed in a simple analytical form, where the velocity along the flow ($v_l$) increases versus the length along the flow ($l$). This is parameterized in \cite{Sim08}, as per their equation~1, where:-

\begin{equation}
v_l =  v_0 + (v_{\infty} – v_0) \left(1 - \frac{R_v}{R_v + l}\right)^{\beta}. 
\end{equation}

Here $v_0$ is the initial velocity at launch, while $R_{\rm v}$ is the radius at which the acceleration starts to occur. In the models adopted here, computed in \cite{Matzeu22} (see their Table 1), $R_v=R_{\rm min}$, i.e. the wind accelerates from the launch point, $v_0=0$ (the initial outwards velocity is deemed negligible) and $\beta=1$ (the velocity power-law index). Thus, with the increasing length ($l$) along the flow, the velocity quickly tends to the terminal velocity, 
$v_{\infty}$ 
 
Furthermore, the wind will not just be launched at a single specific radius, but over multiple radii on the disk, producing a range of terminal velocities originating at different launch radii (between $R_{\rm min}$ and $R_{\rm max}$). 
The terminal velocities realized in the wind models are determined via the launch radius and the terminal velocity parameter ($f_{\rm v}$), where:-
\begin{equation}
v_{\infty} /c = f_{\rm v} \sqrt{2/R}, 
\end{equation}
and $R$ is the launch radius in gravitational units. 
In the spectral fitting, the terminal velocity is adjusted by varying the $f_{\rm v}$ parameter, where $f_{\rm v}$ physically corresponds to the scaling factor between the 
escape velocity at a radius $R$ on the disk and the final terminal velocity. The effect of gravity upon the terminal velocity is also accounted for in the diskwind model, by extrapolation back 
to the launch point of a wind streamline. Thus for $f_v=1$, the wind is launched reaching exactly the escape velocity from the system.

The range in terminal velocities produced by the wind is thus governed by the range in launch radii as well as the $f_{\rm v}$ parameter. 
For example, in the case of $f_{\rm v}=1.5$, then from equation~2, the terminal velocities attained in the DW thick grid range from $v_{\infty}/c=0.21-0.375$, where the fastest 
velocity is launched at the innermost launch radius of $R_{\rm min}=32R_{\rm g}$ and the slowest (outermost radius) at $R_{\max}=96R_{\rm g}$. In contrast the DW thin grid covers a much smaller 
range of terminal velocities over the narrower range of launch radii from $R_{\rm min}=32R_{\rm g}$ to $R_{\max}=48R_{\rm g}$; e.g. for $f_{\rm v}=1.5$, then $v_{\infty}/c=0.31-0.375$. 
Thus the thicker the streamline is in terms of $R_{\rm max}/R_{\rm min}$, the greater the range of velocities realized by the wind. 
As a result, the DW thick grid will tend to produce broader absorption profiles due to a larger velocity shear, as the sightline through the wind 
can intercept a wider range of terminal velocities compared to the DW thin case. 
The free parameters in the model are then:-
\begin{itemize}
\item The terminal velocity parameter, $f_{\rm v}$, which covers the range from $f_{\rm v}=0.25-2.0$ and determines the range of terminal velocities computed. 
\item The inclination angle $\mu = \cos\theta$, where $\theta$ is measured with respect to the polar axis and the minimum wind opening angle ($\theta_{\rm min}$) is 
set at $45\degg$ (see Figure~\ref{fig:schematic}). 
\item The mass outflow rate normalized to the Eddington rate, where $\dot{M}=\dot{M}_{\rm out}/\dot{M}_{\rm Edd}$. 
\item The ionizing X-ray luminosity ($L_{\rm X}$), which is the percentage of the 2--10\,keV luminosity to the Eddington luminosity (i.e. \% $L_{2-10{\rm keV}}/L_{\rm Edd}$). 
\item The photon index of the input X-ray continuum, which is set to that of the hard X-ray power-law (e.g. $\Gamma=2.2$ for \pg). As noted in \cite{Matzeu22}, a steeper $\Gamma$ gives 
stronger absorption lines, due to the weaker ionizing hard X-ray continuum. 
\end{itemize}
Note as both the mass outflow rate and ionizing luminosity are in Eddington units and the wind radius is in gravitational units, the parameters are invariant to the black hole mass. 


\begin{figure}
\begin{center}
\hspace*{-0.8cm}
\rotatebox{-90}{\includegraphics[height=8.5cm]{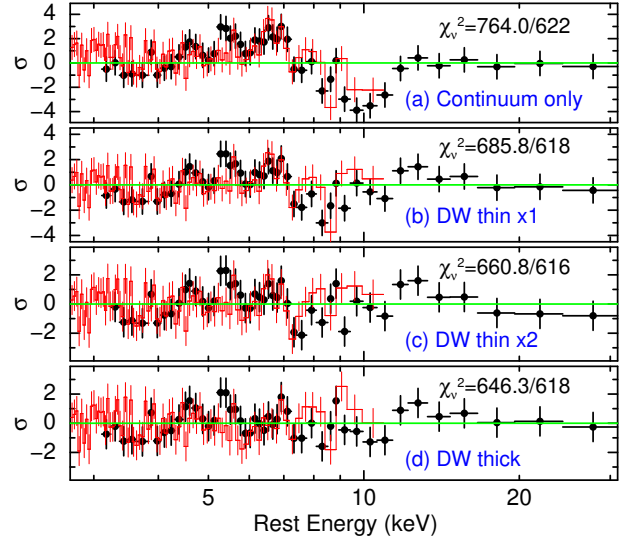}}
\end{center}
\vspace*{-0.3cm}
\caption{Residuals, in units of (data--model)/error, to the mean 2023 spectrum, plotted over the iron K band. Panel (a) shows the residuals to the baseline continuum (no wind), 
panel (b) shows the case of a single DW thin grid, panel (c) corresponds to two DW thin zones and panel (d) corresponds to the best-fit scenario of a single DW thick zone. In each case, the 
parameters are reported in Table~\ref{tab:diskwind}. A geometrically thick disk wind (model (d)) best accounts for the Fe K profile in \pg.}
\label{fig:residuals}
\end{figure}

The above disk wind grids were applied to the mean 2023 spectrum over the 0.3--30\,keV band. 
The same two component baseline continuum was applied (model (a)), where the residuals over the iron K band reveals the wind profile of \pg\ in emission and 
absorption (Figure~\ref{fig:residuals}, panel (a), $\chinu=764.0/622$). 
Initially a single fast DW thin zone is added to the baseline model (model (b), see Table~\ref{tab:diskwind} for parameters). However, while this is able to account for the spectral residuals between 10--12\,keV, at the blue-wards end of the profile, it does not account for the absorption profile between 8--10\,keV, while some of emission remains un-modeled between 6--7 keV (see Figure~\ref{fig:residuals}, panel (b), 
$\chinu=685.8/618$). 
Essentially, the absorption profile predicted by thin wind is not broad enough to account for the whole profile, as the range of terminal velocities achieved is insufficient. A second DW thin zone was then added to the model (model (c), Table~\ref{tab:diskwind}) 
and at a lower terminal velocity in order to account for the remaining residuals. This improves the fit statistic further (Figure~\ref{fig:residuals}, panel (c), $\chinu=660.8/616$), modeling most of the lower energy residuals, although some remain over the 7--10\,keV band. Essentially this solution requires two distinct wind zones, where the fastest zone has a maximum terminal velocity of $v_{\infty}/c=-0.39\pm0.03$, modeling the blue-wards extent of the profile, and a slower zone where the maximum velocity attained is $v_{\infty}/c=-0.23\pm0.03$. The parameters of the two velocity wind are reported in Table~\ref{tab:diskwind} (models (b) and (c) for the two zones). Note the parameters of the faster zone do not change significantly (within errors) upon the addition of the second slower zone. 

\begin{deluxetable}{lc}
\tablecaption{Results of Spectral Fitting to Mean 2023 Spectrum.}
\tablewidth{240pt}
\tablehead{\colhead{Parameter} & \colhead{Value}}
\startdata
(a) Baseline Continuum:-\\
Photon Index, $\Gamma$ & $2.20\pm0.03$ \\
Power-Law Normalization ($N_{\rm PL}^{a}$) & $1.11\pm0.04$ \\
Cross Normalization ({\it NuSTAR}/EPIC-pn) & $1.14\pm0.02$ \\
Seed Photon Temperature, $T_{0} ({\rm keV})$ & $0.074\pm0.004$ \\
Comptonization Temperature, $kT ({\rm keV})$ & $0.40^{+0.17}_{-0.09}$\\
Optical Depth ($\tau$) & $8.3^{+1.9}_{-2.0}$\\
Soft X-ray flux ($F_{\rm 0.3-2\,keV}^b$) & $5.4\pm0.1$\\
Fit Statistic (Baseline only), $\chi^2/\nu$ & $764.0/622$ \\
\hline
(b) Single Thin Disk Wind (Zone 1):-\\
Mass Outflow Rate ($\dot{M}_{\rm out}/\dot{M}_{\rm Edd}$) & $0.67\pm0.10$ \\
Ionizing Luminosity (\% $L_{2-10 {\rm keV}}/L_{\rm Edd}$) & $1.4^{+0.3}_{-0.2}$ \\
Terminal Velocity Parameter ($f_{\rm v}$) & $1.56^{+0.06}_{-0.17}$  \\
Maximum Terminal Velocity ($v_{{\rm max}, \infty}/c$) & $-0.39\pm0.03$ \\
Inclination ($\mu=\cos\theta$) & $0.64\pm0.02$ \\
Fit Statistic (Single Thin Wind), $\chi^2/\nu$ & $685.8/618$ \\
\hline
(c) Additional Thin Disk Wind (Zones 2):-\\
Mass Outflow Rate ($\dot{M}_{\rm out}/\dot{M}_{\rm Edd}$) & $0.48\pm0.09$ \\
Ionizing Luminosity (\% $L_{2-10 {\rm keV}}/L_{\rm Edd}$) & $1.4^t$ \\
Terminal Velocity Parameter ($f_{\rm v}$) & $0.93^{+0.16}_{-0.06}$  \\
Maximum Terminal Velocity ($v_{{\rm max}, \infty}/c$) & $-0.23\pm0.03$  \\
Inclination ($\mu=\cos\theta$) & $0.64^t$ \\
Fit Statistic (Second Thin Wind), $\chi^2/\nu$ & $660.8/616$ \\
\hline
(d) Single Thick Disk Wind:-\\
Mass Outflow Rate ($\dot{M}_{\rm out}/\dot{M}_{\rm Edd}$) & $0.84^{+0.11}_{-0.12}$ \\
Ionizing Luminosity (\% $L_{2-10 {\rm keV}}/L_{\rm Edd}$) & $0.80\pm0.08$ \\
Terminal Velocity Parameter ($f_{\rm v}$) & $1.72\pm0.12$  \\
Maximum Terminal Velocity ($v_{{\rm max}, \infty}/c$) & $-0.43\pm0.03$\\
Minimum Terminal Velocity ($v_{{\rm min}, \infty}/c$) & $-0.25\pm0.02$\\
Inclination ($\mu=\cos\theta$) & $0.58\pm0.03$ \\
2-10\,keV Luminosity, ($L_{\rm 2-10\,keV}^c$) & $2.6\pm0.2$ \\ 
Fit Statistic (Single Thick Wind), $\chi^2/\nu$ & $646.3/618$\\
\enddata
\tablenotetext{a}{Power-law normalization, in units of $\times10^{-3}$\,photons\,cm$^{-2}$\,s$^{-1}$\,keV$^{-1}$ at 1\,keV.}
\tablenotetext{b}{Observed 0.3--2\,keV flux, in units of $\times10^{-12}$\,erg\,cm$^{-2}$\,s$^{-1}$.}
\tablenotetext{c}{Intrinsic 2--10\,keV luminosity, corrected for wind absorption, in units of $\times10^{43}$\,erg\,s$^{-1}$.}
\tablenotetext{t}{Denotes parameter is tied within the model}
\label{tab:diskwind}
\end{deluxetable}

The two zone solution requires two distinct velocity components, yet in this scenario both are assumed to be launched over the same radial range on the disk ($32-48R_{\rm g}$). A more physical 
representation may correspond to the case of a geometrically thick wind, where the dispersion in terminal velocity is naturally produced over a wide range of 
disk radii, as $v_{\infty}\propto R^{-1/2}$ and $R=32-96R_{\rm g}$ in this case. The DW thick grid provides a good account of the whole iron K band profile (Figure~\ref{fig:residuals}, panel (d)) 
and is also statistically the best-fit solution ($\chinu=646.3/618$). The best-fit wind parameters for the DW thick grid are reported in Table~\ref{tab:diskwind} (model (d)). 
Note the inclination angle of $\mu=0.58\pm0.03$ (or $\theta\sim55\pm2\degg$), is consistent 
with the previous modeling of the 2017 and 2022 observations \citep{Reeves23} and is well within the wind opening angle at 45\degg.
In this model, a wide range of terminal velocities are realized; at the inner streamline the maximum terminal velocity is $v_{{\rm max}, \infty}=-0.43\pm0.03c$, while for the outer streamline the minimum terminal velocity is $v_{{\rm min}, \infty}=-0.25\pm0.02c$. The mean terminal velocity, averaged over all launch radii, is $<v_{\infty}/c>\,=-0.34\pm0.03$. This is in agreement with the outflow velocity obtained from the simple \textsc{xstar} 
absorber.

\begin{figure}
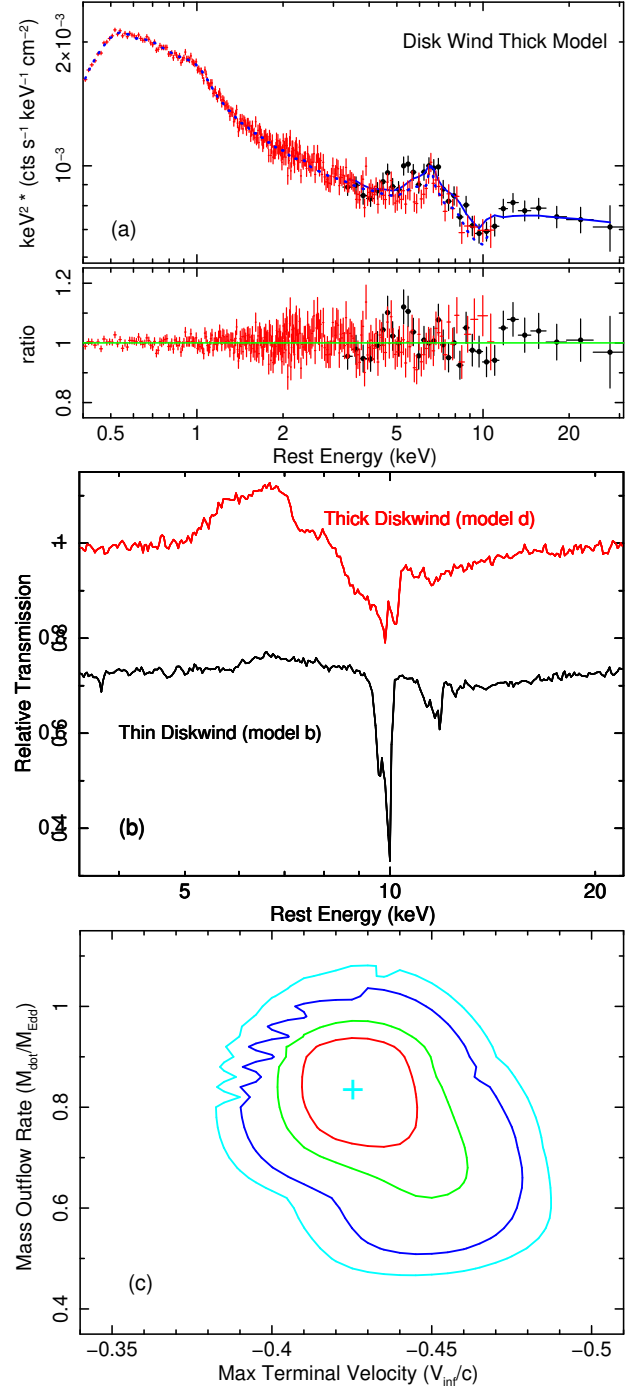

\begin{center}
\hspace*{-0.8cm}
\rotatebox{-90}{\includegraphics[height=8.5cm]{fig7a.eps}}
\hspace*{-0.8cm}
\rotatebox{-90}{\includegraphics[height=8.5cm]{fig7b.eps}}
\hspace*{-0.8cm}
\rotatebox{-90}{\includegraphics[height=8.5cm]{fig7c.eps}}
\end{center}
\vspace*{-0.3cm}
\caption{Results of the disk wind modeling. Panel (a) shows the best-fit DW thick model fitted to the mean 2023 spectrum. Here, the model (folded through the instrumental responses) is shown as a solid blue line fitted to the {\it NuSTAR} data from 3--30\,keV and as a dotted blue line to the \xmm\ spectrum from 0.3--10\,keV. Note the lack of strong soft X-ray absorption. Panel (b) shows the theoretical Fe K profiles predicted from the wind for the DW thick case (red; Table~\ref{tab:diskwind}, model d) versus the DW thin case (black; Table~\ref{tab:diskwind}, model b). A y-axis offset has been applied for clarity. 
The DW thick model produces a broader absorption line profile due to its larger velocity shear, as well as stronger Fe K emission. Panel (c) shows the confidence contours of the mass outflow rate in Eddington units versus the maximum terminal velocity attained by the wind for model (d). The contours represent the 
68, 90, 99 and 99.9\% confidence levels for two interesting parameters. The mass outflow rate in \pg\ appears to be at nearly the Eddington rate, with a maximum terminal velocity of $v_{{\rm max}, \infty}/c=-0.43\pm0.03$. Note the saw-tooth behavior of the outer contours is due to the finite grid resolution.}
\label{fig:diskwind}
\end{figure}

As a result of the wider velocity dispersion within the thick wind, it is able to model the breadth of the absorption profile, unlike the case of a single DW thin zone. 
This is illustrated in Figure~\ref{fig:diskwind}, where panel (a) shows the spectrum fitted with the DW thick model and panel (b) shows the resulting Fe K profile from the wind. 
For the latter plot, the DW thick model (model d) is compared to the single DW thin model (model b). It is apparent that the DW thick solution predicts a substantially  
broader profile compared to the DW thin case, where the absorption trough is notably narrower. In addition, the DW thick model also provides a better description of the Fe K emission from the wind.  

Interestingly, the mass outflow rate (normalized to Eddington) for the DW thick model is $\dot{M}=0.84^{+0.10}_{-0.12}$, which is close to Eddington limit. Indeed the confidence contours of $\dot{M}$ versus $v_{{\rm max}, \infty}$ are well constrained (see Figure~\ref{fig:diskwind}, panel (c)) in this scenario. The mass outflow rate is also much higher than what was determined from the 2022 {\it NuSTAR} observation of \pg. In particular, during the last 60\,ks of that observation (slice B), a broad absorption trough was observed centered near 9\,keV and the mass outflow rate was inferred to be $\dot{M}=0.23\pm0.06$ \citep{Reeves23}. The likely reason for this lower value is the adopted wind geometry, where only the thin disk wind model was considered by \citet{Reeves23} to model the 2022 profile. Instead, the geometrically thicker wind inferred from the higher quality 2023 data naturally requires more mass to achieve a similar line of sight opacity. 
A comparison of the 2023 versus the 2022 epochs with the DW thick model will be considered in Section~5. 

The ionizing X-ray luminosity incident upon the thick wind is found to be $L_{\rm X}=0.80\pm0.08$\%, equivalent to the 2--10\,keV luminosity being 0.8\% of Eddington. 
This level of X-ray luminosity is 
high enough such that the most dominant ionic species is H-like iron and little absorption is predicted at soft X-ray energies. This can also be seen from the best-fit spectrum in Figure~\ref{fig:diskwind} (panel (a)), where the wind model is featureless at soft X-ray energies and no significant residuals are present.
As a consistency check, the above inferred ionizing luminosity can also be compared with the energetics of the AGN. The observed (absorption corrected) 2--10\,keV luminosity  
is $L_{\rm 2-10\,keV}=2.6\pm0.2\times10^{43}$\,erg\,s$^{-1}$ (see Table~\ref{tab:diskwind}). 
From the reverberation black hole mass of $M_{\rm BH}=1.01^{+0.38}_{-0.23}\times10^{7}$\,M$_{\odot}$ \citep{Hu21}, the Eddington luminosity 
is $L_{\rm Edd}=1.3^{+0.5}_{-0.3}\times10^{45}$\,erg\,s$^{-1}$. Based upon these observables, the expected ratio of the 2--10\,keV to Eddington luminosity is $2.0\pm0.6$\%. Within the uncertainties, this is comparable (within a factor of two) to the values inferred from the disk wind modeling; here $L_{\rm X}=0.80\pm0.8$\% for the DW thick case and $L_{\rm X}=1.4^{+0.3}_{-0.2}$\% for the DW thin case. This suggests the disk wind is just slightly under-ionized when compared to the observed 2--10\,keV luminosity. 

\begin{figure*}
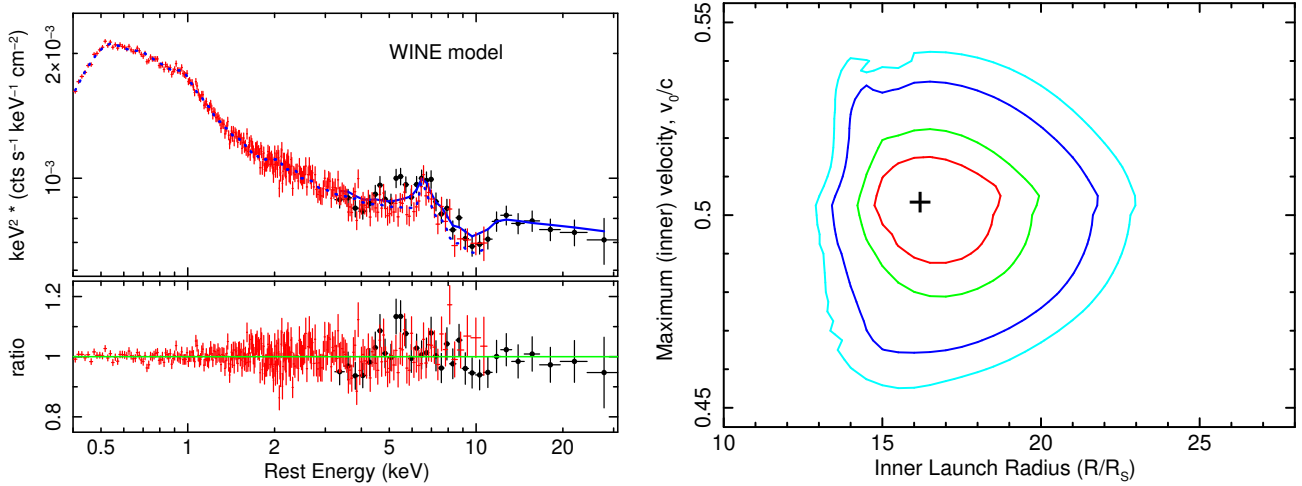

\begin{center}
\hspace*{-0.8cm}
\rotatebox{-90}{\includegraphics[width=6.5cm]{fig8a.eps}}
\vspace*{-0.1cm}
\rotatebox{-90}{\includegraphics[width=6.5cm]{fig8b.eps}}
\end{center}
\vspace*{-0.3cm}
\caption{Results of modeling the mean 2023 \pg\ spectrum with the relativistic \textsc{wine} wind model \citep{Luminari18}. The left panel shows the best fit fluxed spectrum, 
where the folded counts spectrum has been multiplied twice by energy. The \textsc{wine} model is shown as a blue line (solid for {\it NuSTAR}, dotted for EPIC-pn) and reproduces well 
the iron K absorption and emission profile, as is seen by the data/model residuals. The right panel shows the 68\%, 90\%, 99\% and 99.9\% confidence intervals of the maximum wind velocity ($v_0/c$, at the innermost radius) versus 
the inner launch radius ($R_0$), where the latter is in units of the Schwarzschild radius ($R_{\rm s}$). In the best-fit scenario, the wind is launched from a radius of $R_0=16R_{\rm s}$ at a maximum velocity of $0.5c$.}
\label{fig:wine}
\end{figure*}

\subsection{Modeling with the Relativistic WINE model}

Here we apply the \textsc{wine} model of \cite{Luminari18} to provide an alternative parameterization of the wind. The \textsc{wine} model was originally developed to model the 
emission profiles resulting from disk winds, motivated by the P-Cygni profile observed in PDS\,456 \citep{Nardini15}. The model has since been extended to self-consistently calculate the wind absorption and has been applied to the wind in the 2017 \xmm\ observation of \pg\ \citep{Laurenti21} and to the variable wind of NGC\,2992 \citep{Luminari23}. A forthcoming detailed description of the updated model can be found in \citet{Luminari24}.

In summary, the wind geometry is a conical outflow (see Figure~1, \citealt{Luminari18, Luminari24}), with a half-opening angle 
of $\theta_{\rm out}$ with respect to the polar axis. It is launched at an inner radius of $R_0$ (in Schwarzschild units of $R_{\rm s}=2GM/c^2$), with an initial (innermost) ionization parameter of 
$\xi_{0}$. The wind absorption is then integrated over a succession of thin shells versus radius as calculated by \textsc{xstar}, while the wind emission is integrated over all solid angles and radii. Special relativistic effects are accounted for as described in \citet{Luminari20, Luminari21}, including the relativistic de-boosting 
of the continuum as seen by gas expanding radially outwards. The latter effect is important in \pg\ given the large velocities inferred from the iron K absorption profile. 
In order to calculate the resultant line profiles, a wind density profile of $n(r) = n_0 (R/R_0)^{-1}$ is adopted along with a radial velocity profile of $v(r) = v_0 (R/R_0)^{-\frac{1}{2}}$; here $n_0$ and $v_0$ are the inner number density and velocity at the initial launch radius $R_0$. Note the choice of radial profiles is consistent with a momentum conserving wind (e.g. 
\citealt{FGQ12}), while the velocity profile describes a ballistic trajectory de-accelerating under gravity.   

To fit the \pg\ spectrum, the absorption and emission spectra from the \textsc{wine} model were calculated in the form of two \textsc{xspec} multiplicative tables, which modify the same baseline continuum as described earlier. 
The output parameters for spectral fitting are the inner ionization ($\log\xi_0$), inner velocity ($v_0$), inner radius ($R_0$), column density ($N_{\rm H}$) and half opening angle ($\theta_{\rm out}$). 
The first three parameters are assumed to be equal in both absorption and emission, except for the column density which may differ along the line of sight versus (on average) across the whole wind. 
The opening angle is only relevant to the wind emission, as it is calculated over all angles. 
Note that the inclination between the line of sight and the wind symmetry axis is fixed to 0 degrees, since it cannot be constrained by the data, unlike the diskwind model which adopts a bi-conical geometry with a minimum 
opening angle of $45\degg$ (Figure~\ref{fig:schematic}). In WINE, this angle corresponds to a disc observed face-on through a conical, polar wind.  

\begin{deluxetable}{lc}
\tabletypesize{\small}
\tablecaption{Results of fitting the \textsc{wine} model to the 2023 spectrum.}
\tablewidth{240pt}
\tablehead{\colhead{Parameter} & \colhead{Value}}
\startdata
Fitted Parameters:-\\
\hline
Absorption column, $N_{\rm H, abs} /10^{24}\,{\rm cm}^{-2}$ & $0.57\pm0.06$ \\
Ionization, $\log(\xi_{0}/{\rm ergs\,cm\,s^{-1}})$ & $5.58\pm0.04$\\
Inner velocity, $v_0/c$ & $0.50\pm0.02$\\
Inner radius, $R_0/R_{\rm s}$ & $16.3^{+2.6}_{-1.8}$\\
Emission column, $N_{\rm H, emiss} /10^{24}\,{\rm cm}^{-2}$ & $1.7\pm0.3$\\
Half opening angle, $\theta_{\rm out}$ & $>72$\degg\\
Photon Index, $\Gamma$ & $2.14\pm0.02$\\
Fit statistic, $\chi^2/\nu$ & $656.3/618$\\
\hline
Derived Parameters:-\\
\hline
De-boosted luminosity, $L^{\prime}_{\rm ion} / 10^{43}\,{\rm ergs\,s^{-1}}$ & $2.0\pm0.2$\\
Inner density, $n_{0}/ 10^{10}\,{\rm cm}^{-3} $ & $2.1^{+0.9}_{-0.7}$\\ 
Covering fraction, $f_{\rm cov}$ & $>0.69$\\
Mass outflow rate, $\dot{M}_{\rm out} / {\rm M}_{\odot}\,{\rm yr}^{-1}$ & $0.14^{+0.05}_{-0.04}$$^a$\\
Normalized outflow rate, $\dot{M}_{\rm out} / \dot{M}_{\rm Edd}$ & $0.61^{+0.23}_{-0.17}$$^a$\\
\enddata
\tablenotetext{a}{The minimum mass outflow rate, corresponding to $f_{\rm cov}=0.69$.}
\label{tab:wine}
\end{deluxetable}

The \textsc{wine} model produces a good fit to the \pg\ spectrum and in particular over the Fe K band (see Figure~\ref{fig:wine}), with a similar overall fit statistic as per the DW thick case in Section~4.1, with $\chinu=656.3/618$. The fit parameters of the \textsc{wine} model are listed in Table~\ref{tab:wine}, where the maximum wind launch velocity is $v_0/c=-0.50\pm0.02$, just slightly higher than the equivalent maximum terminal velocity obtained by the DW thick model. Note that in the best-fit solution, the minimum wind velocity is $v_{\rm min}=-0.21c$, while the mass averaged velocity is $v_{\rm avg}=-0.325c$. This range of velocity is consistent with what was derived from 
the DW thick grid (model (d), Table~\ref{tab:diskwind}). 

The launch radius is a variable parameter in the \textsc{wine} model and the best-fit value is 
$R_0=16.3^{+2.6}_{-1.8}R_{\rm s}$. This is consistent with the inner launch radius (of $32R_{\rm g}$ or $16R_{\rm s}$) assumed in the disk wind models in Section~4.1 and 
demonstrates that such fast winds are likely to be launched from the innermost regions of the accretion disk. The absorption column density is $N_{\rm H, abs}=0.57\pm0.06\times10^{24}$\,cm$^{-2}$, while the emitter column is higher with $N_{\rm H, emiss}=1.7\pm0.3\times10^{24}$\,cm$^{-2}$. This may occur if the wind is non-uniform, 
for instance if the column density is higher at larger $\theta$ towards the equatorial direction; this is the case for the diskwind models (higher $N_{\rm H}$ at larger $\theta$). 
Finally, from the wind emission, the opening angle is constrained to be $\theta_{\rm out}>72\degg$, which corresponds to a lower limit on the wind geometric covering fraction of $f_{\rm cov} = 1 - \cos(\theta_{\rm out}) > 0.69$. Indeed the larger the opening angle is, the stronger the wind emission will be. 

\begin{figure*}
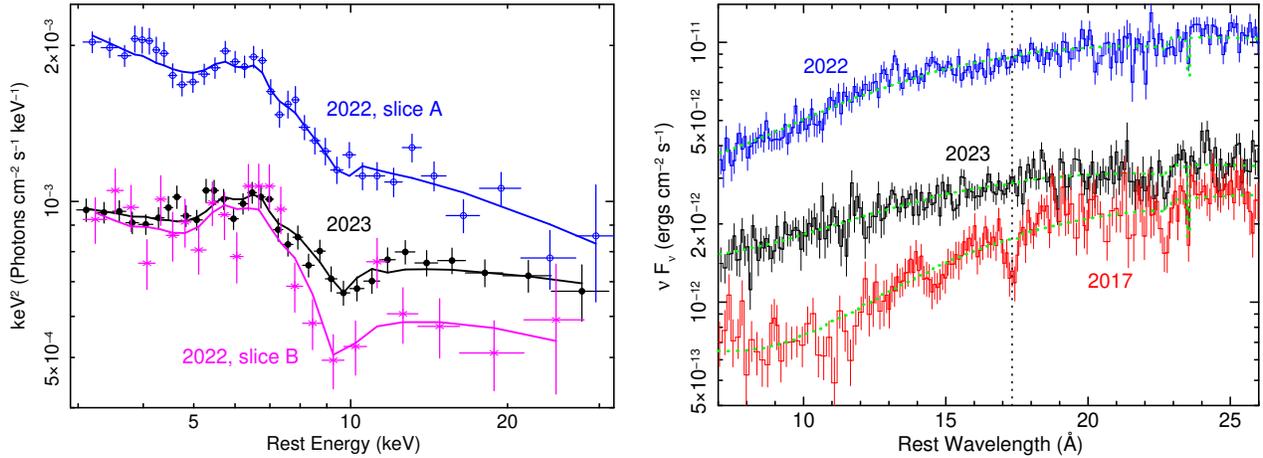

\begin{center}
\hspace*{-0.8cm}
\rotatebox{-90}{\includegraphics[height=8.5cm]{fig9a.eps}}
\rotatebox{-90}{\includegraphics[height=8.5cm]{fig9b.eps}}
\end{center}
\vspace*{-0.3cm}
\caption{
Comparison of the \pg\ spectra over different epochs from 2017--23. The left panel shows the the mean 2023 versus the 2022\,A (slice A) and 2022\,B (slice B) {\it NuSTAR} spectra, fitted with the thick diskwind model in each case. 
The spectral differences between each epoch can be accounted for via changes in ionization through the variations in X-ray luminosity, as well as in photon index. 
Note there was no {\it NuSTAR} observation in 2017. 
The right panel shows the comparison of the 2017, 2022 and 2023 RGS spectra. Significant soft X-ray wind absorption was only detected in the low flux 2017 spectrum, in the form of a 
blueshifted O\,\textsc{viii} Ly-$\alpha$ line, as marked by the dotted vertical line. }
\label{fig:epochs}
\end{figure*}

The innermost density at the launching point of wind is given by the definition of the ionization parameter, where $\xi_0 = L^{\prime}_{\rm ion} / n_0 R_0^2$; here $L^{\prime}_{\rm ion}$ is the 1--1000 Rydberg ionizing luminosity corrected for the de-boosting of the continuum radiation. Following \citet{Luminari20,Luminari23}, the de-boosted luminosity is given by 
$L^{\prime}_{\rm ion} = (\frac{1 - v}{1+v})^{\frac{2 + \Gamma}{2}} L_{\rm ion}$, where $\Gamma$ is the photon index and this also accounts for the redshift of the continuum as seen by the wind. 
Extrapolation of the best-fit baseline continuum model from 1--1000\,Rydberg gives $L_{\rm ion}=2.0\pm0.2\times10^{44}$\,erg\,s$^{-1}$, while from the above the continuum is de-boosted by a factor of $\times10$ for $v_0/c=0.5$ and $\Gamma\approx2.2$. Thus the resultant de-boosted luminosity is $L^{\prime}_{\rm ion} = 2.0\pm0.2\times10^{43}$\,erg\,s$^{-1}$ and the  
innermost density is $n_0=2.1^{+0.9}_{-0.7}\times10^{10}$\,cm$^{-3}$. 

From this and the best fit parameters in Table~\ref{tab:wine}, the mass outflow rate is calculated as per \cite{Laurenti21}:-
    \begin{equation}
          \Dot{M}_\mathrm{out}= 2\,\mu m_\mathrm{p} \int_0^{2 \pi } \int_0^{\theta_\mathrm{out}} \int_{r_0}^{r_1} n(r)\,v(r)\,r \sin\theta\ dr\, d\theta\, d\phi, 
   \end{equation}
    where $\mu, m_\mathrm{p}$ are the mean atomic mass per proton and the proton mass, respectively. The wind outer radius, $r_1$, is calculated analytically from the integral of the density to 
    calculate the absorption column density, while the factor of $\times2$ 
    arises from summing both hemispheres of the wind. From this integral and the radial velocity and density profiles as above, the 
    mass outflow rate is determined to be $\dot{M}_{\rm out} = 0.21^{+0.08}_{-0.06}\,f_{\rm cov}\, {\rm M}_{\odot}\,{\rm yr}^{-1}$. 
    The lower limit of $\theta_{\rm out}>72$\degg\ then yields $f_{\rm cov}>0.69$. 
    Thus the minimum mass outflow rate is $\dot{M}_{\rm out} = 0.14^{+0.05}_{-0.04}\, {\rm M}_{\odot}\,{\rm yr}^{-1}$. For an accretion disk efficiency of $\eta=0.1$ and the black hole mass of \pg\ \citep{Hu21}, this corresponds to a normalized rate of $\dot{M}_{\rm out}/\dot{M}_{\rm Edd} = 0.61^{+0.23}_{-0.17}$ (or $\dot{M}_{\rm out}/\dot{M}_{\rm Edd}=0.91^{+0.35}_{-0.26}$ for $f_{\rm cov}=1$). This is consistent with the mass outflow rate inferred by the disk wind model in Section~4.1.   

\section{Wind Variability}

\begin{deluxetable*}{lcccc}
\tablecaption{Results of Spectral Fitting to OBS 1--4.}
\tablewidth{480pt}
\tablehead{\colhead{Parameter} & \colhead{OBS\,1} & \colhead{OBS\,2b} & \colhead{OBS\,3} & \colhead{OBS\,4}}
\startdata
Photon Index, $\Gamma$ & $2.18\pm0.03$ & $2.11\pm0.04$ & $2.15\pm0.04$ & $2.22\pm0.04$ \\
Power-Law Normalization ($N_{\rm PL}^{a}$) & $1.28\pm0.06$ & $1.24\pm0.08$ & $1.07\pm0.07$ & $1.38\pm0.09$\\
2--10\,keV flux ($F_{\rm 2-10\,keV}^b$) & 2.22 & 2.17 & 1.82 & 2.16 \\
Mass Outflow Rate ($\dot{M}_{\rm out}/\dot{M}_{\rm Edd}$) & $0.73\pm0.19$ & $1.20^{+0.15}_{-0.20}$ & $1.05^{+0.18}_{-0.20}$ & $0.90^{+0.13}_{-0.20}$\\
Ionizing Luminosity (\% $L_{2-10 {\rm keV}}/L_{\rm Edd}$) & $0.80\pm0.13$ & $0.80^t$ & $0.80^t$ & $0.80^t$\\
Maximum Terminal Velocity ($v_{{\rm max}, \infty}/c$) & $-0.46\pm0.03$ & $-0.48\pm0.04$ & $-0.44\pm0.03$ & $-0.43\pm0.04$\\
Minimum Terminal Velocity ($v_{{\rm max}, \infty}/c$) & $-0.27\pm0.02$ & $-0.28\pm0.03$ & $-0.25\pm0.02$ & $-0.25\pm0.03$\\
Inclination ($\mu=\cos\theta$) & $0.58\pm0.04$ & $0.58^t$ & $0.58^t$ & $0.58^t$ \\
Fit Statistic ($\chi^2/\nu$) & $776.9/706$ & $523.9/527$ & $588.5/582$ & $453.7/442$\\
\enddata
\tablenotetext{a}{Power-law normalization ($\times10^{-3}$\,photons\,cm$^{-2}$\,s$^{-1}$\,keV$^{-1}$ at 1\,keV).}
\tablenotetext{b}{Observed 2--10\,keV flux, in units of $\times10^{-12}$\,erg\,cm$^{-2}$\,s$^{-1}$.}
\tablenotetext{t}{Denotes parameter is tied within the model}
\label{tab:4epochs}
\end{deluxetable*}

To test for any wind variability, the DW thick model in Section~4.1 was applied to the four individual 2023 sequences (OBS\,1, 2b, 3, 4), including both the \xmm\ and {\it NuSTAR} spectra 
over the 0.3--30\,keV band. Note the short (dipping) OBS\,2a segment was not included, as its net exposure (6\,ks with EPIC-pn) is too short to derive any useful wind constraints, 
although overall its spectrum is harder with $\Gamma=1.93\pm0.07$. 
The wind parameters in each case are reported in Table~\ref{tab:4epochs}. The results are very similar to the mean spectrum, with no significant variation in the maximum terminal velocity of the 
wind; i.e. from $v_{{\rm max}, \infty}/c=-0.43\pm0.04$ (OBS\,4) to $v_{{\rm max}, \infty}/c=-0.48\pm0.04$ (OBS\,2b). Neither is there any strong variability in the X-ray flux across the epochs. 
The mass outflow rate also remains steady and close to Eddington across the 
four epochs, ranging from $\dot{M}=0.73\pm0.19$ (OBS\,1) to $\dot{M}=1.20^{+0.15}_{-0.20}$ (OBS\,2b). This implies there is no significant intrinsic wind variability across the 40\,day campaign. 

We also compare the properties of the wind in the 2023 mean spectrum with those in the 2022 {\it NuSTAR} observations \citep{Reeves23}. Here we apply the DW thick model to both epochs for consistency. As described in \citet{Reeves23}, the 2022 {\it NuSTAR} observation was split into two slices (slice A and slice B) either side of a factor of $\times2$ drop in flux. The slice A spectrum covered the first 190\,ks in duration of the 2022 {\it NuSTAR} observation, where the source is brightest ($F_{2-10\,{\rm keV}}=4.7\times10^{-12}$\,ergs\,cm$^{-2}$\,s$^{-1}$) and the spectrum was relatively featureless. 
Then in the last 60\,ks of the observation, the flux dropped by a factor of two, reaching a similar level to what is observed in the mean 2023 observation and where a strong absorption trough emerged between 8--12\,keV. 

\begin{deluxetable}{lccc}
\tablecaption{Comparison between the 2023 and 2022 {\it NuSTAR} spectra.}
\tablewidth{240pt}
\tablehead{\colhead{Parameter} & \colhead{2023} & \colhead{2022A} & \colhead{2022B}}
\startdata
$\Gamma$ & $2.16\pm0.03$ & $2.42\pm0.04$ & $2.31\pm0.09$\\
$N_{\rm PL}^{a}$ & $1.29\pm0.08$ & $3.67\pm0.22$ & $1.53\pm0.23$\\
$F_{2-10 {\rm keV}}^b$ & $2.38\pm0.15$ & $4.71\pm0.28$ & $2.25\pm0.34$ \\
$\dot{M}_{\rm out}/\dot{M}_{\rm Edd}$ & $0.91\pm0.22$ & $0.79^{+0.21}_{-0.19}$ & $0.92^{+0.32}_{-0.31}$\\
\%$L_{\rm 2-10 keV}/L_{\rm Edd}$ & $0.58^{+0.13}_{-0.11}$ & $1.16^t$ & $0.58^t$\\
$v_{{\rm max}, \infty}/c$ & $-0.42\pm0.02$ & $-0.42^t$ & $0.42^t$ \\
$\mu=\cos\theta$ & $0.61\pm0.03$ & $0.61^t$ & $0.61^t$\\
$\chi^2/\nu$ & $55.8/62$ & $103.0/104$ & $26.3/21$\\
\enddata
\tablenotetext{a}{Power-law normalization ($\times10^{-3}$\,photons\,cm$^{-2}$\,s$^{-1}$\,keV$^{-1}$).}
\tablenotetext{b}{Flux in units of $\times10^{-12}$\,erg\,cm$^{-2}$\,s$^{-1}$.}
\tablenotetext{t}{Denotes parameter is tied within the model}
\label{tab:nustar}
\end{deluxetable}

The three {\it NuSTAR} spectra (2022\,A, 2022\,B and 2023) were then fitted  with the DW thick grid (model (d)), adopting a simple power-law continuum covering the 3--30\,keV hard X-ray band. These are shown in Figure~\ref{fig:epochs} (left panel), while the results are reported in Table~\ref{tab:nustar}. Given the factor of two difference in luminosity between the brighter 2022\,A spectrum versus the 
2022\,B and 2023 spectra, the ionizing luminosity ($L_{\rm X}$) was set to be a factor of two higher for the former case. 
Similar to what was found above, the mass outflow rate is consistent between all three epochs and thus any difference in the opacity of the wind at iron K simply arises from the higher luminosity of the 2022\,A spectrum, which serves to increase the wind ionization. Thus across the 2022 and 2023 epochs, no intrinsic variations in the wind parameters appear to be observed and they are consistent with a constant terminal velocity and mass outflow rate. The spectral changes can mainly be explained by variations in the X-ray luminosity, along with changes in the photon index.

Finally the soft X-ray wind properties are compared from the respective \xmm\ RGS epochs. The mean 2023 RGS spectrum versus the 2022 and 2017 epochs  
is shown in Figure~\ref{fig:epochs} (right panel). The 2023 RGS spectrum is virtually featureless, with no strong emission or absorption within the $3\sigma$ level, 
and it can be fitted with the baseline continuum model of Section~4 with an acceptable fit statistic ($\chinu=242.8/218$). This is in contrast with the 2017 low X-ray flux epoch, where significant soft X-ray wind features were observed from O\,\textsc{viii} in particular; see \citet{Kosec20, Reeves23}. The vertical line on Figure~\ref{fig:epochs} marks the position of the strong and broadened O\,\textsc{viii} Ly$\alpha$ absorption line in the 2017 spectrum, occurring at $\lambda=17.3\pm0.1$\,\AA\ (rest frame) and blue-shifted by $\sim0.1c$ with respect to the expected wavelength of 18.9\,\AA. 
Indeed, in \citet{Reeves23} a diskwind model with this velocity could account for both the O\,\textsc{viii} and Fe K band absorption in the 2017 epoch. 
In contrast, no such soft X-ray outflow is found in either the 2023 or the bright 2022 epoch, with upper-limits on the equivalent width of O\,\textsc{viii} line of $<1.4$\,eV and 
$<1.8$\,eV for 2023 and 2022 respectively, versus an equivalent width of $5.1\pm1.8$\,eV in 2022. This result is consistent with the lack of soft X-ray absorption from the  
2023 disk wind modeling and suggests the wind is of an overall higher ionization compared to the low flux 2017 epoch. 

\section{Discussion} \label{sec:discussion}

{\it NuSTAR} and \xmm\ observations of \pg\ in 2023 have revealed a remarkable ultra fast outflow, with one of the most extreme blue-shifts measured to date among UFO sources. 
The Fe K band absorption can be parameterized by a simple Gaussian line profile, with a centroid energy of $9.8\pm0.4$\,keV, which, if it is associated to Fe\,\textsc{xxvi} Ly$\alpha$, implies an average blueshift of $v/c=-0.33\pm0.02$ after accounting for the special relativistic correction on the velocity along the line of sight. This is also consistent 
with the mean velocity obtained from the best fitting thick disk wind model, of $<v/c>\,=-0.34\pm0.02$, as is described in Section~4.1 (see Table~\ref{tab:diskwind}, model d). 
The velocity width of the Fe K absorption profile is also pronounced, where $\sigma=1.1^{+0.5}_{-0.4}$\,keV, equivalent to $\sigma_{v}\sim0.1c$. 
This implies a considerable velocity shear, as a result of a range of velocities intercepted along our sightline through the wind. This can be accounted for by the geometrically thick disk wind model, launched over a wide range in radii, where the terminal velocity reaches a maximum value of $v_{{\rm max}, \infty}=-0.43\pm0.03c$ at the inner edge of the wind, with a minimum velocity of 
$v_{{\rm min}, \infty}=-0.25\pm0.02c$ at the outermost radius. The velocity dispersion of $\Delta v=\pm0.1c$ within the wind is consistent with the Gaussian line width as measured above. 
Similar results were also found for the relativistic \textsc{wine} model, where the maximum wind velocity was found to be $v_{0}/c=-0.50\pm0.02$. From the wind emission, the half opening angle of the wind was constrained by \textsc{wine} to be $\theta_{\rm out}>72\degg$ and covering $>2\pi$ steradian solid angle. 

\subsection{Comparison with other winds}

Thus the wind profile in \pg\ resembles the P-Cygni like profile in PDS\,456 \citep{Nardini15}, with broad structure due to emission and absorption from a wide angle outflow. The outflow velocity in \pg\ appears to be at the higher end of what has been measured in the 
prototype UFO in PDS\,456, which typically spans the range between $v/c=0.25-0.35$ \citep{Matzeu17}. Note that in some epochs the fastest outflow component in PDS\,456 can also reach values exceeding $0.4c$ \citep{Reeves18}. A similarly high terminal velocity of $v_{\infty}=-0.38\pm0.02c$ has been reported based upon the disk wind modeling of the similarly luminous QSO, IRAS\,F11119+3257 \citep{Lanzuisi24}. Such velocities may appear to be more commonplace in the most luminous AGN at the peak of the quasar epoch at high redshifts; see \citet{Chartas21} and references therein. 

The velocity measured in \pg\ is much higher than what has been measured in samples of other nearby (e.g. $z<0.1$) AGN 
\citep{Tombesi10, Gofford13}. For illustration, Figure~\ref{fig:ufo} compares the mean disk wind velocity measured in \pg\ with the distributions of velocities obtained from the \xmm\ and {\it Suzaku} UFO 
samples of \cite{Tombesi10} and \cite{Gofford13} respectively. In the \cite{Tombesi10} sample, the mean velocity is $v/c\approx-0.1$, while the distribution of \citet{Gofford13} is skewed 
to lower velocities, with a median velocity of $v/c=-0.056$. There are no AGN across either sample where the outflow velocity exceeds $-0.3c$. This is also similar to what was found in \citet{Igo20}, where the absorption lines were detected through the excess variance spectra of nearby AGN. Likewise, a similar range of velocity was also obtained in the \textsc{subways} sample of \citet{Matzeu23}, which consists of a sample of QSOs at intermediate redshifts ($z=0.1-0.4$). 

Furthermore the equivalent width of the iron K absorption line in \pg, with ${\rm EW}=-435^{+150}_{-190}$\,eV, is much higher than those in other nearby AGN; for instance in the \citet{Gofford13} sample the highest equivalent width is 150\,eV with a mean of 40\,eV (e.g. see their Figure~7), while similar values were also reported by \citet{Tombesi10}. 
Thus among these local AGN, \pg\ has one of the highest reported outflow velocities, while the equivalent width is substantially higher. This implies a 
more powerful wind compared to the predominantly broad lined Seyfert 1 galaxies studied by \citet{Tombesi10} and \citet{Gofford13}. 
In this regard, \pg\ resembles the UFOs reported in other NLS1s, with velocities at the higher range of the AGN distribution from $0.2-0.3c$ and where the absorption features have higher optical depths; e.g. IRAS 13224$-$3809 \citep{Parker17, 
Pinto18, ChartasCanas18}, 1H\,0707$-$495 \citep{Hagino16, Kosec18}, \zw\ \citep{ReevesBraito19}, WKK\,4438 \citep{Jiang18}. 

It appears plausible that the high accretion rates in NLS1s are more conducive for driving a more powerful wind, as they accrete at high Eddington ratios, have strong UV emission and 
generally have steep X-ray spectra with weak hard X-ray emission \citep{GP19}. The latter effect is important, as steep, hard X-ray weak SEDs will tend to produce deeper X-ray wind features due to the gas being less ionized. This was illustrated in the disk wind simulations of \citet{Matzeu22} (see their Figure~3), where the Fe K absorption profile becomes progressively stronger 
as the power-law photon index increases from $\Gamma=1.6$ to $\Gamma=2.4$, with the equivalent widths increasing to up to hundreds of eV in the steepest sources. 
It is also consistent with the observed properties 
of \pg, which has an intrinsically steep hard X-ray spectrum ($\Gamma=2.2-2.4$) and a relatively low inferred ratio of X-ray to bolometric luminosity; e.g. $L_{\rm 2-10 keV}/L_{\rm Edd}\approx1$\%. 

\begin{figure}
\begin{center}
\rotatebox{-90}{\includegraphics[height=8.5cm]{fig10.eps}}
\end{center}
\vspace*{-0.3cm}
\caption{The distribution of wind velocities obtained from two samples of nearby ultra-fast outflows \citep{Tombesi10, Gofford13}, versus the wind velocity in \pg\ observed in 2023. 
The mean velocity in \citet{Tombesi10} is close to $-0.1c$, while the median velocity of the \citet{Gofford13} sample is $-0.056c$, both of which are considerably lower than the average 
terminal velocity of $<v/c>=-0.34\pm0.02$ attained by the \pg\ wind as measured by the DW thick model. Note an even higher value of $v_{\infty}/c=-0.43\pm0.03$ was found for the maximum terminal velocity. Neither of the \cite{Tombesi10} and \cite{Gofford13} samples contain a wind exceeding $>0.3c$ in comparison.}
\label{fig:ufo}
\end{figure}

\subsection{The Variable Wind of \pg}

The velocities of UFOs have been shown to vary in several AGN. Some of the most notable cases reported to date occur in: 
APM 08279$+$5255 \citep{SaezChartas11}, PDS\,456 \citep{Matzeu17}, IRAS\,13224$-$3809 \citep{ChartasCanas18} and \mcg\ \citep{Braito21, Braito22}.
\pg\ also appears to exhibit substantial variability in its terminal velocity across epochs, which may vary by up to a factor of four.  
\xmm\ observations of \pg\ in 2017 caught the AGN in a low flux state and where the iron K absorption trough appeared at a much lower 
centroid energy of 7.5\,keV \citep{Kosec20, Laurenti21, Reeves23} and implying a mean wind velocity of $\approx0.1c$ \citep{Reeves23}. 
In contrast, in the brighter 2022 and 2023 observations, the absorption trough is measured above 9\,keV, while all of these 
later brighter epochs are consistent with the thick disk wind attaining a maximum terminal velocity of $\approx0.4c$. The higher velocities in the 2022--23 epochs 
versus 2017 might be connected to the higher X-ray flux in the later observations. 

The 2017 epoch of \pg\ may also be unusual, as it occurred during a pronounced X-ray dipping 
period. This was observed from the {\it Swift} monitoring during this period \citep{Laurenti21}, where the X-ray flux dropped by up to an order of magnitude compared to the 
optical and UV flux. This decrease may have had a substantial effect on the wind properties, with the wind being substantially slower and also of lower ionization. 
The latter effect is illustrated by the multi-epoch comparison of the RGS spectra (Figure~\ref{fig:epochs}, right panel), where a deep trough was present in the 2017 spectrum due 
O\,\textsc{viii} Lyman-$\alpha$ at a velocity of $-0.1c$, as reported in \cite{Kosec20} and \cite{Reeves23}, but which is absent during the brighter 2022--23 periods. 
Further monitoring with \xmm\ and {\it NuSTAR} will help to reveal whether \pg\ continues to maintain the large velocity measured in 2022--23, or 
whether the AGN wind reacts to any prolonged decreases in luminosity via its ionization and velocity as in the 2017 epoch.

\subsection{Kinematics and Wind Driving Mechanisms}

Here we estimate the kinematics of the disk wind in \pg\ from the 2023 observations.  The mass outflow rates and terminal velocities derived from the disk wind modeling in Section~4.1 (DW thick case) are adopted  (Table~\ref{tab:diskwind}, model (d)). Here, the resultant mass outflow rate ($\dot{M}$) is expressed in Eddington units. To provide a more conservative estimate of the energetics, the average terminal velocity attained by the wind is adopted (i.e. $v/c=-0.34\pm0.02$), rather than the maximum terminal 
value which is achieved at the innermost edge of the streamline. 
The kinetic power in Eddington units ($\dot{E}$) is subsequently:-

\begin{equation}
\dot{E}=\frac{L_{\rm kin}}{L_{\rm Edd}} = \frac{1}{\eta} (\gamma -1) \dot{M} 
\end{equation}

\noindent where $\gamma$ is the Lorentz factor and $\eta$ is the accretion efficiency (and where $\eta=0.1$ is adopted here). 
The corresponding wind momentum thrust in Eddington units is subsequently:-

\begin{equation}
\dot{p}=\frac{\dot{p_{\rm out}}}{\dot{p}_{\rm Edd}} = \frac{1}{\eta} \dot{M} \frac{v}{c}.
\end{equation}

\begin{deluxetable}{lc}
\tablecaption{Outflow Energetics of \pg\ in 2023.}
\tablewidth{240pt}
\tablehead{\colhead{Parameter} & \colhead{Value}}
\startdata
$<v_{\infty}/c>$$^a$ & $-0.34\pm0.02$\\
$\dot{M}^{b}$ & $0.84\pm0.12$ ($0.19^{+0.11}_{-0.06}M_{\odot}$\,yr$^{-1}$) \\
$\dot{E}^c$ & $0.53\pm0.09$ ($6.8^{+4.0}_{-2.4}\times10^{44}$\,erg\,s$^{-1}$)\\
$\dot{p}^{d}$  & $3.0\pm0.5$ ($1.3^{+0.8}_{-0.5}\times10^{35}$\,dyne)\\
\enddata
\tablenotetext{a}{Mean terminal velocity of the wind} 
\tablenotetext{b}{Mass outflow rate in Eddington units, as per Table~2 (absolute value in parenthesis).} 
\tablenotetext{c}{Outflow kinetic power in Eddington units.}
\tablenotetext{d}{Outflow momentum rate in Eddington units.}
\label{tab:energetics}
\end{deluxetable}

The outflow energetics of \pg\ in 2023 are listed in Table~\ref{tab:energetics}, while consistent results are also obtained with the \textsc{wine} model. Note the absolute values are included in parenthesis, which take into account the uncertainties on the 
black hole mass measured by \citet{Hu21}. As a result of the large wind velocity and the high mass outflow rate determined by the disk wind modeling, as well as the wide angle nature of the wind, the kinetic power during the 2023 observations reaches about 50\% of the Eddington luminosity. This is at least an order of magnitude higher than what is postulated to be significant in terms of mechanical feedback in the host galaxy \citep{HopkinsElvis10}. Furthermore, the normalized momentum rate (compared to Eddington) also exceeds unity, i.e. $\dot{p}_{\rm wind}/\dot{p}_{\rm Edd}\approx3$. In contrast a ratio of about one is predicted for black hole winds, e.g. \citet{KP03}, arising from the average single electron scattering limit for the radiation field in a moderately Compton thick wind ($N_{\rm H}\sim10^{24}$\,cm$^{-2}$). 
This is also generally higher than the observed values derived from ultra fast outflow samples, e.g. \citet{Tombesi13, Gofford15}. 

We also compare the mass outflow rate computed by the diskwind (DW thick case) and \textsc{wine} models with that obtained by the more simple \textsc{xstar} model in Section~4. Here the general form of the mass outflow rate is adopted, according to \citet{Tombesi13} and \citet{Gofford15}, where:-

\begin{equation}
\dot{M} = 4\pi f_{\rm cov} \mu m_{\rm p} N_{\rm H} v_{\rm out} R. 
\end{equation}
 
Here $\mu m_{\rm p}$ is the average baryonic particle mass ($\mu=1.27$ for cosmic abundances) and $f_{\rm cov}$ is the geometrical covering, while $N_{\rm H}=6.2^{+1.5}_{-2.1}\times10^{23}$\,cm$^{-2}$ and $v/c = -0.34\pm0.02$ from the \textsc{xstar} fits. 
A covering of $f_{\rm cov}=0.7$ is adopted to provide a like for like comparison with the \textsc{wine} model, while likewise a launch radius of $R=16 R_{\rm S}$ is used. 
For these values, then subsequently $\dot{M}_{\rm out} = 0.09\pm0.03$\,M$_{\odot}$\,yr$^{-1}$, or equivalently 
$\dot{M}_{\rm out}/\dot{M}_{\rm Edd}=0.40\pm0.12$ (for $\eta=0.1$). The latter value is similar to, but somewhat lower than the values computed by the diskwind model (Table~2, DW Thick case) and \textsc{wine} (Table~3). 
On the other hand, the \textsc{xstar} value may be somewhat underestimated, as it does not account for special relativistic effects and the de-boosting of the continuum as seen by the wind, 
as is calculated in \cite{Luminari20}. 
 
Overall it may be difficult to accelerate and sustain such a fast powerful wind by continuum (i.e. Thomson) radiation pressure alone, even for an Eddington limited AGN. 
The maximum possible wind velocity in the radiative case will also be limited by special relativistic effects. Here, the ability of radiation alone to drive a wind velocity as high as  
$-0.4c$ will be further restricted, due to the non-negligible de-boosting of the radiation field as received by the outflowing gas \citep{Luminari20, Luminari21}. 

In contrast, magneto hydrodynamical (or MHD) winds are not limited by such effects and potentially can drive iron K winds from the innermost accretion disk to velocities as high as 
$-0.6c$ \citep{Fukumura10}. As illustrated in Figure~3 in \citet{Fukumura10}, the most favorable conditions occur in AGN with steep X-ray spectra (either in $\alpha_{\rm ox}$ or $\Gamma$).  
The relative paucity of hard X-ray photons then prevents over ionization of either Fe\,\textsc{xxv} or Fe\,\textsc{xxvi} in the innermost wind regions, where the fastest part of the wind is launched.  
Such models are able to reproduce the fast wind in PDS\,456, attaining a maximum wind velocity of up to $-0.39c$ in that case \citep{Fukumura18}. 
The large velocity width of the absorption profile in \pg\ is reminiscent of the profiles predicted by MHD winds, especially for steep $\Gamma$ sources due to the more favorable ionization 
conditions \citep{Fukumura18, Fukumura22}. This is as a result of the large range in radii over which the wind is launched on the disk, where $v_{\infty} \propto R^{-1/2}$. 
In both the MHD and radiation cases \citep{Matzeu22}, a high accretion rate with respect to Eddington is also beneficial, as this will increase the outflow density, further lowering the wind ionization. 
Thus in the high accretion rate NLS1s, both the properties of the incident radiation field and MHD processes can play an important part in accelerating and sustaining disk winds to such high velocities. 

However the large outflow rates may be at least in part mitigated by a lower geometrical covering of the gas. 
The broad emission component determines the overall extent of the wind. However if part of the emission profile is 
formed by a relativistic accretion disk line, e.g. see \citet{Parker22} or \citet{Middei23}, then the restrictions on the covering fraction could be relaxed. 
This would help to reconcile the derived mass outflow rate with UV line driven wind simulations. For example in \citet{Nomura20}, 
the mass outflow rates can approach 50\% of Eddington as the Eddington limit is reached; such rates are largely consistent with what is observed here.   
As demonstrated by \citet{Mizumoto21}, UV line driven winds can be inhomogeneous and produce a wide range in observed outflow velocities as a result of the in situ acceleration of the wind. 
The strong UV to soft X-ray excesses in NLS1s such as \pg, in contrast to AGN with higher masses and low Eddington ratios, may also enhance a line driven wind. 
Furthermore, in the mildly super Eddington regime, powerful, yet clumpy winds can even be formed by continuum (i.e. Thomson) radiation pressure alone \cite{Takeuchi13}. 
The properties of these winds, whether they are clumpy and/or inhomogeneous, will be soon revealed by the {\it Resolve} calorimeter on {\it XRISM} \citep{Tashiro20}, 
which thanks to its high spectral resolution will accurately probe the velocity field of the outflowing gas.



\section{Acknowledgements}
JR and VB acknowledge financial support through NASA grants 80NSSC22K0474 and 
80NSSC23K1467. D.P acknowledges financial support from the CNES french spatial agency. 
A. Luminari acknowledges support from the HORIZON-2020 grant “Integrated Activities for the High Energy Astrophysics Domain" (AHEAD-2020), G.A. 871158). 
SH acknowledges support from the Science and Technologies Facilities Council (STFC) through the studentship grant ST/V506643/1. 
Based on observations obtained with {\it XMM-Newton}, an ESA science mission with instruments and contributions directly funded by ESA Member States and NASA and from the {\it NuSTAR} mission, a project led by the California Institute of Technology, managed by the Jet Propulsion Laboratory, and funded by NASA

\end{document}